\documentclass[prd,twocolumn,tightenlines,showpacs,nofootinbib,amsfonts,amssymb,amsmath]{revtex4-1}
\usepackage{graphicx}
\begin{document}
\newcommand{\etal}{{\it et al.}}
\newcommand{\bx}{{\bf x}}
\newcommand{\bn}{{\bf n}}
\newcommand{\bk}{{\bf k}}
\newcommand{\dd}{{\rm d}}
\newcommand{\dslash}{D\!\!\!\!/}
\def\ga{\mathrel{\raise.3ex\hbox{$>$\kern-.75em\lower1ex\hbox{$\sim$}}}}
\def\la{\mathrel{\raise.3ex\hbox{$<$\kern-.75em\lower1ex\hbox{$\sim$}}}}
\def\beq{\begin{equation}}
\def\eeq{\end{equation}}

\leftline{UMN--TH--3209/13}

\vskip-2cm
\title{Anisotropy in solid inflation}

\author{Nicola Bartolo$^{1,2}$, Sabino Matarrese$^{1,2}$, Marco Peloso$^{2,3}$, 
Angelo Ricciardone$^{1,2}$}
\affiliation{
${^1}$ Dipartimento di Fisica e Astronomia ÒG. GalileiÓ, \\
Universit\`a degli Studi di Padova, I-35131 Padova (Italy) \\
${^2}$ INFN, Sezione di Padova, I-35131 Padova  (Italy)\\
${^3}$ School of Physics and Astronomy,
University of Minnesota, Minneapolis, 55455 (USA)\\
}
\vspace*{2cm}

\begin{abstract}
In the model of solid / elastic  inflation,   inflation  is driven by a source that has the field theoretical description of a solid.
To allow for prolonged  slow roll  inflation, the solid needs to be extremely insensitive to the spatial expansion. We point out that, because of  this property, the solid is also rather inefficient in erasing  anisotropic deformations of the geometry. This allows for a prolonged inflationary anisotropic solution, providing the first  example  with standard gravity and  scalar fields only which evades the conditions of  the so called cosmic no-hair conjecture. We compute the curvature perturbations on the anisotropic solution, and the corresponding phenomenological bound on the anisotropy.  Finally, we discuss the  analogy between this model and the $f \left( \phi \right) F^2$ model, which also allows for  anisotropic inflation thanks to a suitable coupling between the inflaton $\phi$ and a vector field. We remark that the bispectrum of the curvature perturbations in solid inflation is enhanced in the  squeezed limit and presents a nontrivial angular dependence, as had previously been found for the 
 $f \left( \phi \right) F^2$ model.
\end{abstract}
 \date{June 2013}
 \maketitle

\section{Introduction}

The CMB data strongly support the inflationary framework and allow to rule out several specific inflationary models  \cite{Ade:2013uln}. 
Still, a large number of models remains compatible with the data. It proves useful to classify and study  them  in terms of an effective field theory description of inflation \cite{Cheung:2007st,Weinberg:2008hq}, where the behavior of the perturbations and the possible signatures can be understood in terms of symmetries and symmetry breaking. For instance, several models of inflation are characterized by a shift symmetry $\phi \rightarrow \phi + C$ (where $\phi $ is the inflaton, and $C$ a constant), which protects the required flatness of the inflaton potential against radiative corrections (for a recent review see \cite{Pajer:2013fsa}). In the limit of exact shift symmetry the potential coincides with a cosmological constant, and the spacetime geometry is the de Sitter one. A small and controlled breaking of the symmetry ensures a slow roll inflaton evolution, which breaks time translational invariance. This set-up has one scalar perturbation, which can be identified as the Goldstone boson of this broken symmetry.

While the requirement that inflation ends demands that time translation invariance is broken, and, more in general, cosmology 
studies  time-evolving backgrounds, the vast majority of the models assumes invariance under spatial translations, in agreement with the observed  homogeneity and isotropy of the universe at large scales.  In fact, one of the many features of inflation is that it can dynamically lead to the observed homogeneity and isotropy \cite{Linde:2005ht}, which are otherwise hard to achieve starting from more general initial conditions  \cite{Collins:1972tf}. 

The simplest way to enforce isotropy and homogeneity is to assume that only spin zero fields are dynamically relevant during inflation, and that their vacuum expectation value (vev) is independent of the spatial coordinates. These assumptions characterize the vast majority of the  models of inflation. However, in principle, one could imagine that different sources are present, which individually break the invariance under spatial transformations, but that their combined effect  - due to some underlying symmetry - preserves the background isotropy and homogeneity. Due to this different symmetry breaking pattern, such a possibility could result in specific phenomenological signatures that are not  obtained in the more conventional inflationary models. 

Isotropy with spin one sources can be achieved through (i) a triplet of orthogonal vectors with equal vev  \cite{triad}, (ii) a large number $N \gg 1$ of randomly oriented vectors \cite{Golovnev:2008cf}  - resulting in a  ${\rm O } \left( 1 / \sqrt{N} \right) \ll 1$ anisotropy -  or massive vectors oscillating about the minimum of the potential  \cite{massive-V} - resulting in an effective isotropic equation of state once averaged over the oscillations. 
 
Homogeneity and isotropy with spin zero fields  with a spatially-dependent vev can be achieved with a triplet of scalars with  \cite{ArmendarizPicon:2007nr,Endlich:2012pz}
\begin{equation}
\langle \phi^i \rangle = x^i \;,
\label{vev-phi} 
\end{equation}
where $i=1,2,3$. In particular, ref. \cite{Endlich:2012pz} dubbed this model {\it Solid Inflation}. It is assumed in  \cite{Endlich:2012pz} that the medium driving inflation can be coarse-grained at the level of fundamental cells, and that only the position of these cells is relevant for inflationary cosmology. The three scalars   $\phi^i \left( t , \vec{x} \right)$ can be viewed as the three coordinates that provide the position at the time $t$ of the cell element that at the time $t=0$ was at position $\vec{x}$. Therefore, the vevs (\ref{vev-phi}) characterize a medium at rest in comoving coordinates.  The properties of the solid are defined through a lagrangian, which is a functional of  $\phi^i$. As we describe below, (i) only derivatives of the scalars enter in the lagrangian, so that the vev (\ref{vev-phi}) can be compatible with homogeneity of the background solution, and (ii)  only SO(3) invariant combination of the derivatives enter in the lagrangian, so that  the vev (\ref{vev-phi}) can be  compatible with isotropy. As discussed in  \cite{Endlich:2012pz},  this description provides a complementary formulation of \cite{Gruzinov:2004ty}, which also suggested a coarse-grained description of the inflationary medium dubbed {\it Elastic Inflation}. 

By construction, the background evolution of solid inflation is isotropic, and the power spectrum of the scalar perturbations  is statistically isotropic  \cite{Gruzinov:2004ty,Endlich:2012pz}.  However, the bispectrum presents a characteristic shape not encountered in previous models of scalar field inflation  \cite{Endlich:2012pz}. Specifically, it is enhanced in the squeezed limit as the local template, but it manifests a nontrivial dependence on the angle between the small and large momentum in the correlator. It turns useful to adopt the parametrization \cite{Shiraishi:2013vja} 
\begin{equation}
 B_\zeta \left( k_1 , k_2 , k_3 \right) = \sum_L c_L P_L \left( {\hat k}_1 \cdot {\hat k}_2 \right) P_\zeta \left( k_1 \right) P_\zeta \left( k_2 \right) +  2 \, {\rm perm.} \,,
\label{squeezed-B}
 \end{equation}
 where $P_\zeta$ and $B_\zeta$ are, respectively, the power spectrum and bispectrum of    the curvature perturbation in the uniform density gauge and $P_L$ are the Legendre polynomials. The local template is characterized by $c_i = \frac{6}{5} f_{\rm NL} \delta_{i0}$. More in general, typical models of scalar field inflation are characterized by $c_i = 0$, for $i \neq 0$ in the squeezed limit. The reason for this is the following
  \cite{Babich:2004gb,Lewis:2011au}:   in the squeezed limit $k_3 \ll k_{1,2}$, the long wavelength mode modulates the two short wavelength modes when they leave the horizon. From the point of view of the short wavelength modes,  the long wavelength mode can be accurately described by a mean and   a gradient. The gradient defines a local basis for a quadrupolar dependence of the small-scale power, thus in principle contributing to the $c_2$ coefficient above. However, the gradient vanishes in the long wavelength limit $k_3 \rightarrow 0$. 
  
On the contrary, the nonvanishing scalar vevs (\ref{vev-phi}) provide a directionality modulation of the bispectrum that does not vanish in the squeezed limit, and the bispectrum of solid inflation is dominated by the $c_2$ term in the squeezed limit  \cite{Endlich:2012pz}. Quite interestingly, a nontrivial angular dependence in that limit had previously been obtained in \cite{Barnaby:2012tk} and further studied in \cite{Bartolo:2012sd,Funakoshi:2012ym,Shiraishi:2013vja,Abolhasani:2013zya,Biagetti:2013qqa,Fujita:2013qxa} 
~\footnote{Ref. \cite{Lyth:2013sha} rederived the results of  \cite{Bartolo:2012sd}, claiming that their rederivation uses only the classical mode functions, and it is therefore ``simpler and more complete'' than the computation of  \cite{Bartolo:2012sd}. The rederivation is not more complete,  since, by  admission, it disregards the contribution from the modes in the quantum regime.  We argue that it is also not simpler, since also the results of  \cite{Barnaby:2012tk,Bartolo:2012sd} are due to the classical super-horizon contribution,  as  repeatedly stressed  in \cite{Bartolo:2012sd}.} in the model $f\left( \phi \right) F^2$, where $\phi$ is the inflaton and $F^2$ the square of a vector field  strength $F_{\mu \nu}$, in the case in which $f \left( \phi \right)$ is chosen so to produce a scale invariant spectrum for the vector field. In this  case, the nontrivial angular dependence of the bispectrum is due to the fact that a homogeneous vector breaks isotropy locally, and so the anisotropic modulation survives also in the $k_3 \rightarrow 0 $ limit.~\footnote{Notice that in the $f\left( \phi \right) F^2$ model with a non-vanishing vev of the vector field, a bispectrum that breaks statistical isotropy is generated, and its angle-average does 
assume the form~(\ref{squeezed-B}). A statistically isotropic bispectrum is obtained from a triplet of orthogonal vectors of equal magnitude.} This has further nontrivial consequences, as we discuss in the Conclusions. For the  $f\left( \phi \right) F^2$ model, $c_2 = c_0 /2$, while all other $c_i$ coefficients vanish.

Motivated by the above models, the Planck collaboration \cite{Ade:2013ydc} has  constrained the first coefficients of the series (\ref{squeezed-B}), as $c_0 = 3.24 \pm 6.96$,  $c_1 = 11.0 \pm 113$, and $c_2 = 3.8 \pm 27.8$ (all at $68\%$ CL), with error bars in agreement with the forecasts of  \cite{Shiraishi:2013vja}. 

The $f \left( \phi \right) F^2$ mechanism is constructed to generate and sustain a nontrivial vector field in cosmology (see \cite{Maleknejad:2012fw} for a recent review and for a more extended list of relevant works).  A vector field with standard   ${\cal L} = - \frac{1}{4} F^2$ lagrangian  is   conformally coupled to a FRW background, and so its fluctuations are not excited by the expansion of the universe. Moreover, if a vector vev is present as an initial condition, it is rapidly diluted away by the expansion of the universe. Therefore, any signature associated to the vector - including the angular dependence in (\ref{squeezed-B}) - would be negligible in this case. Several models have been proposed for which a classical vector vev is not diluted by the expansion. 

Several of them break the gauge invariance associated with the vector field. Such models 
 are characterized by (i) a suitable vector potential $V \left( A^2 \right)$ \cite{Ford:1989me}, (ii) a specific coupling to the scalar curvature ${\cal L} \supset \frac{1}{12} R A^2$ \cite{Turner:1987bw,Golovnev:2008cf,Dimopoulos:2008yv}, or (iii) a lagrange multiplier $\lambda$ that enforces a fixed norm for the vector, ${\cal L} \supset \lambda \left( A^2 - v^2 \right)^2$  \cite{Ackerman:2007nb}.  Due to the broken gauge invariance, the vector field has also a longitudinal mode. This mode   turns out to be a ghost \cite{hcp} in all the above models. On the contrary, the  $f \left( \phi \right) F^2$ mechanism preserves gauge invariance, and it is therefore stable \cite{Himmetoglu:2009mk}. A suitable choice of $f \left( \phi \right)$ can result in  frozen and scale invariant super-horizon  perturbations, and in a constant vev, for the magnetic or electric component of the vector field. The first possibility has been suggested as a model for inflationary magnetogenesis \cite{Ratra:1991bn,Martin:2007ue,Giovannini:2009xa} (although this application is highly nontrivial to realize 
\cite{Demozzi:2009fu,Barnaby:2012tk,Fujita:2012rb,Ferreira:2013sqa}), while the second one has been used to obtain a prolonged stage of anisotropic inflationary expansion  \cite{Watanabe:2009ct}.~\footnote{See \cite{aniso-fAA} for models of anisotropic inflation  that employ the idea of  \cite{Watanabe:2009ct}.} 
  
  The $f \left( \phi \right) F^2$ mechanism and solid inflation constitute the two only examples known so far of a primordial bispectrum with a nontrivial angular dependence in the squeezed limit. It is natural to ask whether 
   the two models have other common aspects, and in fact the present investigation originated by an argument 
  that convinced us that the analogy between the models already starts at the background level: 
a remarkable property of the medium of solid inflation is that it is very weakly affected by the huge inflationary expansion. This property, which is completely at odds with that of the solids that we ordinarily deal with, is encoded by an extremely weak dependence of the energy of the solid on its volume. Cosmological perturbations in solid inflation are supported by deformation of this solid - the ``phonons''.~\footnote{See \cite{Matarrese:1984zw} for an early lagrangian formulation of the cosmological medium as a fluid, and for the description of its perturbations in terms of phonons. A different lagrangian formulation of a fluid driving inflation has also been recently studied in  \cite{Arroja:2010wy,Chen:2013kta}.} Stability of these perturbations, and the existence of a weak coupling regime, require that the medium is not only very weakly sensitive to the overall volume expansion, but to all spatial deformations  \cite{Endlich:2012pz}. This naturally led us to conjecture that the solid should be extremely inefficient to respond to anisotropic background deformations, and that, consequently, it should also admit prolonged anisotropic solutions. The computations of the present work show that this is indeed the case.

Specifically, we obtain that the anisotropy is erased on a timescale $\Delta t = {\rm O } \left( \frac{1}{\epsilon H} \right)$, where $H$ is the Hubble rate, and $\epsilon$ the slow roll parameter $\epsilon \equiv - \dot{H} / H^2$ (dot denoting a time derivative). This corresponds to the isotropization rate $\Delta t^{-1} = {\rm O } \left( \epsilon H \right)$. This rate is suppressed with respect to the isotropization rate $\Delta t^{-1} = {\rm O } \left( H \right)$ that is typically encountered in inflationary models  
\cite{Wald:1983ky,Maleknejad:2012as}.  Ref. \cite{Wald:1983ky} showed that a ${\rm O } \left( H \right)$ isotropization rate is the norm for practically all homogeneous and anisotropic backgrounds (with the possible exception of a  Bianchi type-IX geometry) in the presence of a cosmological constant and a fluid that satisfies the  dominant and strong energy conditions. This result is often denoted in the literature as the ``cosmological no-hair conjecture'' (or ``theorem''), as it implies that no information on the anisotropy survives, analogously to what would happen for a black-hole solution. The above cited vector field models are attempts to evade the results of  \cite{Wald:1983ky}, and, as we discussed, only those based on the $f \left( \phi \right) F^2$ represent viable solutions. Besides using vector fields, other works that have attempted to evade  the result of  \cite{Wald:1983ky} involve either  higher order  curvature terms \cite{R2aniso}  or higher forms 
 \cite{Kaloper:1991rw,DiGrezia:2003ug,Ohashi:2013mka}. Ref.  \cite{Ohashi:2013mka} supports the higher form through the same mechanism as  \cite{Watanabe:2009ct}, and it is therefore stable. To our knowledge, a full study of the perturbations for the  proposals  \cite{Kaloper:1991rw,DiGrezia:2003ug} remains to be done. The one we present here is the first counter example of  \cite{Wald:1983ky} with standard gravity and only scalar fields. This counter example has no pathologies: as we show below, the slow isotropization in solid inflation  precisely originates by the demand that the phonons have a well behaved propagation in this unconventional medium.

The work is organized as follows. In Section \ref{sec:model} we present the model of solid / elastic inflation, and its FRW solution,  as formulated in \cite{Endlich:2012pz}. In Section \ref{sec:FRW}  we review the curvature perturbation on the FRW solution,  again mostly  summarizing the original study of \cite{Endlich:2012pz}. In Section \ref{sec:Bianchi-bck} we study the simplest  anisotropic solution in this model and we discuss why the result of \cite{Wald:1983ky} is evaded. In Section  \ref{sec:Bianchi-perts} we study the scalar curvature perturbation on this anisotropic solution and we obtain the corresponding phenomenological limit on the anisotropy.   In the concluding Section  \ref{sec:conclusions} we further   discuss the analogy between solid inflation and the     $f \left( \phi \right) F^2$ model, and we review some interesting open questions.

\section{The model and the FRW background solution}
 \label{sec:model}

The action of solid inflation is  \cite{Endlich:2012pz}
\begin{equation}
S = \int d^4 x \sqrt{-g} \left\{ \frac{M_p^2}{2} R + F \left[ X, Y, Z \right] \right\} \;,
\label{solid}
\end{equation}
where $R$ is the scalar curvature, $M_p$ the (reduced) Planck mass, and $F$ a function that characterizes the solid, as we now discuss.
The solid is divided in several infinitesimal cells. The three scalars   $\phi^i \left( t , \vec{x} \right)$ can be viewed as the three coordinates that provide the position at the time $t$ of the cell element that at the time $t=0$ was at position $\vec{x}$. Therefore, the vevs (\ref{vev-phi}) characterize a medium at rest in comoving coordinates.   To reconcile a homogeneous and isotropic solution with background fields that are ${\bf x}-$dependent, ref.  \cite{Endlich:2012pz}  imposes that 
the function $F$ is invariant under translations $\phi^i \rightarrow \phi^i + C^i$, and   SO(3) rotations, $\phi^i \rightarrow O^i_j \phi^j$, with $C^i, O^i_j$ constant. Specifically, it is assumed that $F$ is a function of SO(3) invariants of
\begin{equation}
B^{ij} \equiv g^{\mu \nu} \partial_\mu \phi^i \partial_\nu \phi^j\, .
\label{Bij}
\end{equation}
Only three independent such invariants, exist, that in \cite{Endlich:2012pz} are chosen as~\footnote{The determinant of $B^{ji}$ can be written as the combination  ${\rm det \; } B = \frac{X^3}{6} \left( 1 - 3 Y + 2 Z \right)$. Since the energy of a perfect fluid is only sensitive to volume deformations, we can regard the special case in which $F$ only depends on this combination as the field theoretical description of a fluid. Such a case was also discussed in  \cite{Endlich:2012pz} and studied in \cite{Ballesteros:2012kv}.}

\begin{equation}
X \equiv {\rm Tr } \,  B =  B^{ii} \;\;\;,\;\;\; Y \equiv \frac{{\rm Tr } \,  \left( B^2 \right)}{\left( {\rm Tr  } \, B \right)^2} \;\;\;,\;\;\;
 Z \equiv \frac{{\rm Tr } \left( B^3 \right)}{\left( {\rm Tr  } \, B  \right)^3} \;.
\end{equation}
As we shall see, the SO(3) invariance in the ``internal $\left\{ \phi^i \right\}$   space'',          
together with the ``diagonal'' vevs (\ref{vev-phi}), allows for an isotropic background solution for the model. However, it is important to stress that this is not the only admissible solution. In fact, in Section  \ref{sec:Bianchi-bck} we will see that the vev (\ref{vev-phi}) is compatible with anisotropic solutions, for which the anisotropy is encoded in different scale factors for the   different spatial directions (a Bianchi-I background). In the reminder of this Section we concentrate on the isotropic background solution
\begin{equation}
d s^2 = - d t^2 + a^2 \left( t \right) d x^i d x^i \,\,,
\label{FRW-line}
\end{equation}
and we also state the conditions for the validity of the theory obtained in \cite{Endlich:2012pz} (which apply to generic backgrounds). 

 The energy momentum tensor obtained from (\ref{solid}) is 
\begin{equation}
T_{\mu \nu} = g_{\mu \nu} F - 2 \partial_\mu \phi^i \partial_\nu \phi^j \, \frac{\partial F}{\partial B^{ij}} \,\,,
\label{Tmunu}
\end{equation}
where,
\begin{eqnarray}
\frac{\partial F}{\partial B^{ij}} & = & \left( F_X - \frac{2 Y}{X} F_Y - 3 \frac{Z}{X} F_Z \right) \delta^{ij} \nonumber\\
& &  + \frac{2 F_Y}{X^2} B^{ij}
+ \frac{3 F_Z}{X^3} B^{ik} B^{kj} \,\,.
\label{FBij}
\end{eqnarray}
On the background (\ref{FRW-line}), the three above invariants have the vevs $\langle X \rangle = \frac{3}{a^2}, \langle Y   \rangle = \frac{1}{3} , \langle Z \rangle = \frac{1}{9} $, and     we obtain $\langle T^\mu_\nu \rangle  = {\rm diag } \left( - \rho , p , p , p \right) $, with
\begin{equation} 
\rho = - F \;\;,\;\; p = F - \frac{2}{a^2} \, F_X
\label{rho-p}
\end{equation}
where the subscript denotes partial derivative.  The background Einstein equations are the   standard ones in terms of the above energy density and pressure
\begin{equation}
3 H^2 = \frac{\rho}{M_p^2} \;\;,\;\;  - 2 \dot{H} - 3 H^2  = \frac{p}{M_p^2}
\label{FRW-eq}
\end{equation}

In  addition, one has the equations obtained by extremizing the action with respect to the scalar fields:
\begin{equation}
\partial_\mu \left[ \sqrt{-g} \frac{\partial F}{\partial \partial_\mu \phi^i } \right] = 
\partial_\mu \left[ \sqrt{-g} \frac{\partial F}{\partial B^{ab}}    \frac{\partial B^{ab}}{\partial \partial_\mu \phi^i } \right] = 0 \,\,.
\label{eom-phi}
\end{equation}

For the above background configuration (\ref{vev-phi}), 
\begin{equation}
 \frac{\partial B^{ab}}{\partial \partial_\mu \phi^i} \Big \vert_{\phi^i = x^i} = \delta_i^a g^{\mu b} +  \delta_i^b g^{\mu a}    \,\,.
\end{equation}
As the indices  $a$ and $b$ only range from $1$ to $3$, as long as the metric is diagonal the expression (\ref{eom-phi}) 
automatically vanishes when $\mu =0$. Moreover, as long as the metric is $\vec{x}$-independent, the expression in square parenthesis is 
also $\vec{x}$-independent. Therefore, the equations (\ref{eom-phi}) are automatically (that is, for any functional form of $F \left[ X, Y, Z \right]$) satisfied by  (\ref{vev-phi}), both on a FRW and on a Bianchi-I background.

Following  \cite{Endlich:2012pz}, we define the slow roll parameters,
\begin{equation}
\epsilon    \equiv  \frac{-\dot{H}}{H^2} =  \frac{X \, F_X}{F  } \;\;,\;\;
\eta    \equiv  \frac{\dot{\epsilon}}{\epsilon H} 
= 2 \left( \epsilon - \frac{X F_{XX} + F_X}{F_X} \right) \,\,,
\label{epsilon-eta}
\end{equation}
and we impose that   $ \epsilon  , \vert \eta \vert \ll 1$, as required for successful inflation.  From eq. (\ref{rho-p}), we see that $F < 0$. We then impose that $\dot{H} < 0$ during inflation, which forces $F_X < 0$.

Let us now discuss the validity of the effective field theory that describes the solid \cite{Endlich:2012pz}.     To do this, it is sufficient to study the perturbations of a solid with $ \vert X F_X \vert \ll \vert F \vert$ on a Minkowski background (as always, this study reproduces the study of cosmological perturbations in the sub-horizon regime  \cite{Endlich:2012pz}). We decompose an arbitrary deformation of the solid (namely, a ``phonon''), $\phi^i = x^i + \pi^i \left( x \right)$ into longitudinal plus transverse one, $\vec{\pi} = \vec{\pi}_L + \vec{\pi}_T$, with, respectively, the properties  $\vec{\nabla} \times \vec{\pi}_L = 0$ and $\vec{\nabla } \cdot \vec{\pi}_T$. The sound speed of such perturbations is~\footnote{The approximation made in the second equation in (\ref{cTcL}) is $\vert X F_{XX} + F_X \vert \ll \vert F_X \vert$, as it is  required to have $\vert \eta \vert \ll 1$ in eq. (\ref{epsilon-eta}). The full expression for the longitudinal sound speed that we use in the following computations is $c_L^2 = \frac{4}{3} c_T^2 - 1 + \frac{2}{3} \epsilon - \frac{1}{3} \eta$.
} \cite{Endlich:2012pz}
\begin{equation}
c_T^2 = 1 + \frac{2}{3} \, \frac{F_Y + F_Z}{F_X X} \;\;,\;\; c_L^2 \simeq \frac{4}{3} c_T^2 - 1 \,\,.
\label{cTcL}
\end{equation}

It is then immediate to verify that the requirements of subluminal propagation of the perturbations ($c_{L,T}^2 < 1$) and of the absence of tachyonic modes 
($c_{L,T}^2 > 0$) are obtained for  \cite{Endlich:2012pz}
\begin{equation}
0 < F_Y + F_Z < \frac{3}{8} \epsilon \, \vert F \vert \,\,.
\label{sound-condition}
\end{equation}
As the ``phonons'' enter derivatively in $F$, their nonlinear interactions necessarily become strong at energies $E $ greater than some scale $ \Lambda $. We need to require that $\Lambda \gg H$, so that there exist a finite window of sub-horizon scales in which the theory 
(\ref{solid}) is weakly coupled. A detailed study performed in  \cite{Endlich:2012pz} shows that this is the case for $\epsilon c_L^3 \gg \left( \frac{H}{M_p} \right)^{2/3}$. This condition can be satisfied at sufficiently  small $H$. This condition, together with (\ref{sound-condition}) ensures that the field theoretical description (\ref{solid}) of the solid is under perturbative control.

To conclude this Section,  we note that  $F_X$ is the only derivative of the function $F$ that  enters in the expression for the pressure, since, by construction, $Y$ and $Z$ are insensitive to the overall spatial volume \cite{Endlich:2012pz}. Therefore, $F_X$ is the only quantity that characterizes the sensitivity of the solid to the volume expansion.   We need to impose that  $\epsilon \ll 1$, or,  equivalently,   that this sensitivity is extremely small. This is not a surprising condition: the source of inflation needs to have an equation of state sufficiently close to that of a cosmological constant, which is, by definition, insensitive to the volume expansion. This property is in complete contrast with that of solids that we ordinary deal with, but nonetheless it is logically conceivable, and it has a perfectly valid field theoretical description  \cite{Endlich:2012pz}.  From the study of the perturbations of such an unusual medium, we learn that also   the combination $F_Y + F_Z$ needs to be small. As we shall see in Section \ref{sec:Bianchi-bck}, this combination,   obtained from  the sound speed of the  phonons, controls the response of the solid to anisotropic deformations.  We therefore learn that the solid not only needs to be    extremely insensitive   to the volume expansion, but also to an anisotropy of the geometry. This   property is    the basis for the prolonged anisotropic inflationary  solution that we obtain in Section \ref{sec:Bianchi-bck}.

\section{ Scalar curvature perturbations on the FRW solution}
 \label{sec:FRW}

In this section we summarize the linearized  study  \cite{Endlich:2012pz}  of the perturbations on the FRW background of solid inflation discussed in the previous Section. We decompose the fields in background plus perturbations
\begin{eqnarray}
\phi^i & = & x^i + \pi^i \left( t , \vec{x} \right) \;\;\;,\;\;\;  \pi^i \left( t , \vec{x} \right) = \frac{\partial_i}{\sqrt{-\partial^2}}  \pi_L
 + \pi_T^i \,\,,  \nonumber\\
g_{00} & = & - 1 - 2  \Phi \left( t , \vec{x} \right) \,\,,  \nonumber\\ 
g_{0i} & = &    B^i \left( t , \vec{x} \right) \;\;\;,\;\;\;  B^i \left( t , \vec{x} \right) = \frac{\partial_i }{\sqrt{-\partial^2}}  B_L+ B_T^i \,\, ,  \nonumber\\
g_{ij} & = & a^2 \left( t \right) \left( \delta_{ij} + h_{ij} \left( t , \vec{x} \right) \right)  \,\,,
\label{perts}
\end{eqnarray}
where   $\pi_T^i$ and  $B_T^i$ are transverse, and $h_{ij}$ is transverse and traceless. 
The perturbations of the metric are classified according to how they transform under spatial rotations. The perturbations $\Phi$ and $B_L$ 
transform as two scalar modes, the perturbations $ B_T^i $ form a vector multiplet (of two degrees of freedom, given the transversality condition), and the perturbations $h_{ij}$ form a tensor multiplet (again of two degrees of freedom). Modes with different transformation properties are decoupled from one another at the linearized level. We note that we have set to zero two scalar modes and one vector mode in $\delta g_{ij}$, leading to the so called spatially flat gauge. This can always be done using infinitesimal coordinate transformations, and actually this fixes completely this gauge freedom (equivalently, one may choose to use gauge invariant combinations of the perturbations \cite{Bardeen:1980kt,Mukhanov:1990me}).

In addition, also the perturbations of the scalar field are separated into a ``longitudinal'' and a ``transverse'' part. The $\pi_T^i$ multiplet is not a vector multiplet under spatial rotations, given that all fields $\phi^i$ are scalar fields, and the index $i$ is in this case just a label for the three fields. However, due to the fact that at the background level $\langle \phi^i \rangle = x^i$, one can verify that, at the linearized level, 
  $\pi_L$ only couples to the scalar modes of the metric, while  $\pi_T^i$ only couples to the vector multiplet of the metric. Therefore,  with an abuse of notation, in the following we refer to $\pi_L$ as a scalar perturbation, and to  $\pi_T^i$ as a vector multiplet.
  
Therefore, after fixing the freedom of infinitesimal coordinate transformation, the system of perturbations has a scalar sector of $3$ degrees of freedom ($\pi_L, \Phi, B_L$), a vector sector of $4$ degrees of freedom ($\pi_T^i, B_T^i$)    and a tensor sector of $2$ degrees of freedom. However, not all these degrees of freedom represent physically propagating independent degrees of freedom. The modes $\Phi, B_L, B_T^i$, that form the  $\delta g_{0 \mu}$ elements enter in the quadratic action of the perturbations  without time derivatives, and are not independent degrees of freedom \cite{Arnowitt:1962hi}. In Fourier space, the equations of motion for these non-dynamical fields are algebraic in them, and can be solved to give the non-dynamical fields as a function of the dynamical fields, without introducing any additional degree of freedom. Therefore the system of physically propagating perturbations of the model consists of one scalar degree of freedom, $\pi_L$, two ``vector'' degrees of freedom, $\pi_T^i$, and two tensor degrees of freedom, $h_{ij}$ (the latter are the two polarizations of the gravitational waves).  For our purposes we are interested only in the scalar sector at the linearized level and we refer the interested reader to \cite{Endlich:2012pz} for a detailed analysis of the vector and tensor  modes at the linearized level, and for the calculation of the three point function of the scalar mode at the non-linear level in a FRW background.

The scalar / vector decomposition appearing in (\ref{perts}) is better understood in Fourier space. We Fourier transform each perturbation 
$\delta \left( t , \vec{x} \right)$ as
\begin{equation}
\delta \left( t, \vec{x} \right) = \int \frac{d^3 k}{\left( 2 \pi \right)^{3/2}} {\rm e}^{i \vec{x} \cdot \vec{k} } \, \delta \left( t , \vec{k} \right) \,\,,
\end{equation}
(we use the same symbol for the mode in real and in Fourier space, as the context always makes manifest which  of the two our following equations refer to). Then, if ${\hat k}_i $ denotes the unit vector  in the direction of the momentum of the mode,   we have $\pi_L = -i {\hat k}_i \cdot \pi^i$, and $\pi_T^i = \pi^i + i  {\hat k}_i \pi_L$ (and identically for $B^i$).  

To study the scalar sector at the linearized level, we expand the action (\ref{solid}) at quadratic order in the Fourier modes of $\pi_L, \Phi,B_L$. The algebraic equations for $\Phi$ and $B_L$ obtained by extremizing this action are, respectively, solved by
\begin{eqnarray}
\Phi & = &   k  \epsilon a^2 H \, \frac{\dot{\pi}_L + \epsilon H \pi_L}{k^2+3 \epsilon a^2 H^2} \,\,, \nonumber\\ 
B_L & = & \epsilon a^2 H \frac{-3 a^2 H \dot{\pi}_L + k^2 \pi_L}{k^2+3 \epsilon a^2 H^2} \,\,.  
\end{eqnarray}

Inserting  these solutions back into the quadratic  action we obtain the free action for the scalar physical degree of freedom
\begin{eqnarray}
S  & = &  \int d t d^3 k 
a^3 M_p^2 \Bigg(  \frac{\epsilon a^2 H^2 k^2}{k^2+3 \epsilon a^2 H^2} 
\left\vert  \dot{ \pi }_L + \epsilon H \pi_L \right\vert^2 \nonumber\\
&& \quad\quad\quad\quad  \quad\quad\quad\quad  \quad\quad
- \epsilon H^2 c_L^2 k^2 \lvert \pi_L \vert^2 \Bigg)  \; ,
  \label{azpi-FRW}
 \end{eqnarray}
in agreement with  \cite{Endlich:2012pz}. From this expression we recognize that the speed of the scalar perturbations is indeed $c_L$ in the flat space-time / sub-horizon regime.

We are interested in the gauge invariant variable $\zeta$, that represents  the curvature perturbation  on 
uniform-density hypersurfaces. In our gauge
\begin{eqnarray}
\zeta \Big\vert_{\delta g_{ij,{\rm scalar}} = 0} & \equiv & - H \frac{ \delta \rho }{ \dot{\rho} } = -  \frac{ k}{3} \pi_L  \,\, ,  \nonumber\\
\label{zeta}
\end{eqnarray}
As shown in  \cite{Endlich:2012pz}, the variable $\zeta$ is continuous if the end of inflation and reheating occur due to a sharp phase transition that modifies $F$.
Therefore, we are interested in the value that $\zeta$ assumes on  super-horizon scales during inflation. The initial condition for $\zeta$ is obtained by computing the canonically normalized  variable, in terms of which the action (\ref{azpi-FRW}) acquires the form
\begin{equation}
S  =  \frac{1}{2} \int d \tau d^3 k \left[ \vert V' \vert^2 - \omega^2 \vert V \vert^2 \right] 
\;\;\Rightarrow\;\; V_{\rm in} = \frac{ {\rm e}^{-i \int^\tau \omega d \tau' + i \phi_0} }{\sqrt{2 \omega}
}  \,\,.
  \label{azV-FRW}
\end{equation}
(the relation between $\zeta$ and $V$ is immediately obtained by comparing the kinetic term of  (\ref{azpi-FRW})  and of   (\ref{azV-FRW})). In this expression, prime denotes derivative with respect to conformal time $\tau$, and $\phi_0$ is an arbitrary unphysical phase.  The initial condition is the so called adiabatic vacuum solution,  set in the deep sub-horizon regime, where the frequency is  adiabatically evolving, $\omega' \ll \omega^2$.

As shown in  \cite{Endlich:2012pz}, it is actually convenient to consider the curvature perturbations ${\cal R}$, that in spatially flat gauge,
is related to $\zeta$ by
\begin{equation}
{\cal R} =  \frac{1}{\epsilon H} \frac{\dot{\zeta} + \epsilon H \zeta}{1+k^2 / \left( 3 a^2 \epsilon H^2 \right)} \;\;,
\label{RZeta}
\end{equation}
since the equation of motion for ${\cal R}$,
\begin{eqnarray}
&& {\cal R}'' + \left( 2 + \eta - 2 s_L \right) a  H  {\cal R}' + k^2 c_L^2 {\cal R} 
 + \Bigg[ 3 \epsilon - 6 s_L + 3 c_L^2 \epsilon \nonumber\\
&& \quad\quad   - \epsilon \left( 2 \epsilon + \eta \right) + 2 s_L \left( 2 \epsilon - \eta \right) + s_\eta \eta \Bigg] a^2 H^2 {\cal R} = 0 \;\;,
\label{eqR}
\end{eqnarray}
is significantly simpler than the one for $\zeta$. In this expression, $s_L \equiv \frac{\dot{c_L}}{c_L H}$   \cite{Endlich:2012pz}, and
 $s_\eta \equiv \frac{\dot{\eta}}{\eta H}$  are slow roll-suppressed quantities. Eq. (\ref{eqR}) is exact, but  the  second line (not explicitly given in
 \cite{Endlich:2012pz}) is second order in slow roll and negligible for all the following considerations. Up to first order in slow roll, the solution is    
\begin{eqnarray}
{\cal R} & = & C \left( \frac{ \tau }{\tau_c}  \right)^{-\alpha}  H_\nu^{(1)} \left( - c_L k \tau \left( 1 + s_{L,c} \right) \right) \;\;, \nonumber\\
\alpha & \equiv & - \frac{1}{2} \left( 3 + 2 \epsilon_c + \eta_c - 2 s_{L,c} \right) \;\;\;, \nonumber\\
\nu & \equiv & \frac{1}{2} \left( 3 + 5 s_{L,c} - 2 c_{L,c}^2 \epsilon_c + \eta_c \right) \;\;, 
\label{solR}
\end{eqnarray}
where $\tau_c$ is some time during inflation, and the  suffix $c$  indicates that the corresponding quantity is evaluated at $\tau_c$. We have already eliminated the solution $\propto  H_\nu^{(2)} $ which approximates to a negative frequency mode in the asymptotic past.

We take the time derivative of eq.  (\ref{RZeta}) and we combine it with  the  equation of motion for $\zeta$ following from (\ref{azpi-FRW}),~\footnote{We also need to use the explicit solutions for the background quantities given in Appendix A of  \cite{Endlich:2012pz}.} 
so to eliminate $\ddot{\zeta}$.  We obtain an equation relating $\dot{\cal R}$, ${\zeta}$, and $\dot{\zeta}$. The system formed by this equation and by  eq.  (\ref{RZeta}) con be formally solved to express $\zeta$ and its derivative in terms of ${\cal R}$ and its derivative.  We then insert the explicit solution (\ref{solR}) and its time derivative into these formal expressions, and obtain
\begin{eqnarray}
\zeta & = & C \left( \frac{\tau}{\tau_c} \right)^{3/2} \left[ 1 + \left( \epsilon_c + \frac{\eta_c}{2} - s_{L,c} \right) \,  {\rm ln } \frac{\tau}{\tau_c} \right] \nonumber\\
&& \quad\quad \quad \times
\left[ - \frac{\epsilon_c}{3} H^{(1)}_\nu \left( Q \right) 
 +  \frac{ k \tau}{3 c_L} \left( 1 - \epsilon_c \right) \,  H^{(1)}_{1+\nu} \left( Q \right) \right] \,\,,  \nonumber\\
\label{explicit-zetasol}
 \end{eqnarray}
where, for brevity,      $Q \equiv  - k \tau c_L \left( 1 + s_{L,c} \right) $. This expression is  valid up to   first order in slow roll.~\footnote{The explicit expression  (\ref{explicit-zetasol}) has not been given in   \cite{Endlich:2012pz}, and  we reported its derivation since some of the subdominant terms in $\zeta$ and $\zeta'$ are needed for the power spectrum computation that we perform in Section  \ref{sec:Bianchi-perts}. We have written this expression in the most compact way; doing so, however, it contains also terms which are second or higher order in slow roll, and which should be disregarded. Such terms do not enter in any of our computations.}

 The coefficient  $C$ can be now set from evaluating the solution (\ref{explicit-zetasol}) at the asymptotic past $\tau_{\rm in} \rightarrow - \infty$, and from the consideration made right after  eq. (\ref{azV-FRW}):
\begin{equation}
 C = - i \sqrt{\frac{\pi}{2}} \frac{\left( -\tau_c \right)^{3/2}   \, c_{L,c} H_c  }{ 2  M_p \sqrt{ \epsilon_c } } 
\label{norma-zeta}
\end{equation}
(up to subdominant slow roll corrections), where the arbitrary phase has been chosen so that $\zeta$ is real and positive in 
the asymptotic past during inflation. In fact,  using (\ref{explicit-zetasol}) and (\ref{norma-zeta}), 
 we can finally write the expression for $\zeta$ in the late time / super-horizon regime ($ - k c_L  \tau \ll 1 $):
\begin{eqnarray}
 \zeta_{\rm late} & \simeq & \frac{ H_c}{2 k^{3/2} c_L^{5/2} M_p \sqrt{\epsilon_c }} 
  \left\{ 1 + \epsilon_c \left[ \left( 1 + c_{L,c}^2 \right) \, {\rm log } \frac{\tau}{\tau_c } + {\rm O } \left( 1 \right) \right] \right\}
  \,\,, \nonumber\\
  \label{zetalarge-gen}
\end{eqnarray}
in agreement with  \cite{Endlich:2012pz}. It is worth pointing out that  the variable $\zeta$ presents a (slow roll suppressed) growth outside the horizon  \cite{Endlich:2012pz}. One of the conditions for the conservation of $\zeta$ on super-horizon scales is that the anisotropic part of the stress-energy tensor vanishes in that regime~\cite{Bardeen:1980kt}. This is the case for  minimally coupled scalar fields with  $\vec{x}-$independent vev \cite{Malik:2008im}. In the present model, however, 
\begin{eqnarray}\label{scalar stress}
\delta T_{ij,\rm{scalar}} & = & a^{2} M_{p}^2\dot{H}\zeta\Bigg[2\left(3-2\epsilon+\eta\right)\delta_{i j} \nonumber\\
& & 
-\left(3+3c_{L}^2-2\epsilon+\eta\right)
\left(3\hat{k}_{i}\hat{k}_{j}-\delta_{i j}\right)\Bigg] \,\,,
\end{eqnarray}	 
where we recall that ${\hat k}_i$ is the unit-vector in the direction of the momentum of the mode.
In the standard case, the anisotropic part can be at best proportional to spatial gradients, and therefore vanishes in the large scale limit. This is not the case in the present model, due to the $\vec{x}-$dependent scalar field vevs. We note that $\delta T_{ij,\rm{scalar}} $ is slow roll suppressed,  which explain why the evolution of $\zeta$ on super-horizon scales  is also slow-roll suppressed.

\section{Prolonged anisotropic background  solution }
 \label{sec:Bianchi-bck} 

Let us now consider  a Bianchi-I background
\begin{eqnarray}
&& d s^2 =  - d t^2 + a^2 \left( t \right) d x^2 + b^2 \left( t \right) \left[ d y^2 + d z^2 \right]    \,\,, \nonumber\\
&& a \equiv {\rm e}^{\alpha - 2 \sigma} \;\;,\;\;
b \equiv {\rm e}^{\alpha +  \sigma}  \,\,,
\label{bianchi}
\end{eqnarray}
where, for simplicity, we have assumed a residual $2$d isotropy in the $y-z$ plane (we expect that dropping this assumption would complicate the algebra, without affecting  the main  physical conclusions of this and of the next Section). We follow the notation of \cite{Watanabe:2009ct} of parametrizing by ${\rm e}^\alpha$ the ``average'' scale factor (the volume scales as $\sqrt{-g} = {\rm e}^{3 \alpha}$) and by  ${\rm e}^\sigma$ the anisotropy. In principle, one could also consider anisotropic vevs for the scalar fields, $\langle \phi^i \rangle = c^i x^i$, with $c^i$ being three different constants. However, starting from such configuration, one can always rescale coordinates so that the relation (\ref{vev-phi}) is maintained, and the line element is still of the form (\ref{bianchi}).

As we discussed after eq.  (\ref{eom-phi}), the scalar fields equations of motion are solved by the ansatz  (\ref{vev-phi}) and (\ref{bianchi}). Let us therefore turn our attention to the Einstein equations ${\rm Eq}^\mu_\nu \equiv G^\mu_\nu - \frac{T^\mu_\nu}{M_p^2}$, 
and, using (\ref{Tmunu}) and (\ref{FBij}), we obtain

\begin{eqnarray}
&&    
 \dot{\alpha}^2 -  \dot{\sigma}^2 + \frac{F}{3 M_p^2} = 0 \,\,,  \nonumber\\
&&             
 \ddot{\alpha} + 3 \dot{\sigma}^2-\frac{  {\rm e}^{4 \sigma} + 2 {\rm e}^{-2 \sigma}}{3 M_p^2} {\rm e}^{-2 \alpha} F_X = 0  \,\,, \nonumber\\
&&         
 \ddot{\sigma} + 3 \dot{\alpha} \dot{\sigma} -  \frac{2}{3} \frac{{\rm e}^{4 \sigma} - {\rm e}^{- 2 \sigma}}{M_p^2} {\rm e}^{-2 \alpha} F_X  \nonumber\\
&& \quad\quad   
 - \frac{4 {\rm e}^{6 \sigma} \left( {\rm e}^{6 \sigma} - 1 \right) F_Y}{\left( {\rm e}^{6 \sigma } + 2 \right)^3 M_p^2}  - \frac{6 {\rm e}^{6 \sigma} \left( {\rm e}^{12 \sigma} - 1 \right) F_Z}{\left( {\rm e}^{6 \sigma } + 2 \right)^4 M_p^2} = 0 \,\,, \nonumber\\
\label{ein-bianchi}
\end{eqnarray}
which correspond, respectively, to the $\frac{{\rm Eq}^0_0}{3} $, $\frac{{\rm Eq}^1_1 + 2 {\rm Eq}^2_2 - 3 {\rm Eq}^0_0 }{6}$, and $\frac{{\rm Eq}^1_1 -  {\rm Eq}^2_2 }{3} $ combinations of the Einstein equations. Due to the background symmetries,  ${\rm Eq}^3_3   = {\rm Eq}^2_2   $, while  ${\rm Eq}^\mu_\nu $ identically vanish for $\mu \neq \nu$. Moreover, the three equations (\ref{ein-bianchi})    are actually not independent, since they are related by a nontrivial Bianchi identity $\left( \frac{d}{d t} + 3 \dot{\alpha} \right)  Eq^0_0 - \left( \dot{\alpha} - 2 \dot{\sigma} \right) Eq^1_1 - 2 \left( \dot{\alpha} + \dot{\sigma} \right) Eq^2_2 = 0$. Therefore, a closed set of sufficient equations for the two scale factors  is obtained by taking for instance the first two, or the first and the third one among (\ref{ein-bianchi}).

The observed statistical isotropy of the CMB constrains the background anisotropy to be small (we quantify this statement in the next Section). Therefore, we restrict the study of the Einstein equations   to the $\sigma \ll 1$ regime. Up to $ {\rm O } \left( \sigma^2 \right) $ corrections, the first two equations in (\ref{ein-bianchi}) reduce to the FRW equations (\ref{FRW-eq}), where $H = \dot{\alpha}$. The third equation gives instead
\begin{equation}
\ddot{\sigma} + 3 \dot{\alpha} \dot{\sigma} - \frac{4 {\rm e}^{-2 \alpha } F_X + \frac{8}{9} \left( F_Y + F_Z \right)}{M_p^2} \sigma + {\rm O } \left( \sigma^2 \right) = 0 \,\,.
\end{equation}
Using (\ref{cTcL}), this equation rewrites
\begin{eqnarray}
&& \ddot{\sigma} + 3 H \dot{\sigma} + 4 \epsilon H^2 c_T^2 \sigma + {\rm O } \left( \sigma^2 \right) = 0 \,\,, \nonumber\\
&& H \equiv \dot{\alpha} \;\;,\;\; \epsilon \equiv - \frac{\dot{H}}{H^2} \,\,.
\label{eq-sigma}
\end{eqnarray}
where we have recalled the definitions of the ``average'' Hubble rate $H$, and of the slow roll parameter $\epsilon$. As ${\rm O } \left( \sigma^2 \right)$ are disregarded in (\ref{eq-sigma}), such quantities can be evaluated from the FRW equations  (\ref{FRW-eq}), disregarding the anisotropy.

In the case of standard scalar field inflation, the normalization of the scale factors is unphysical (for a flat geometry), and the anisotropy is encoded in the ``anisotropic Hubble rate'' $h \equiv \dot{\sigma}$. This quantity obeys the equation $\dot{h} + 3 H h = 0$ (see for instance \cite{Gumrukcuoglu:2007bx}), which corresponds  to the first two terms in (\ref{eq-sigma}). This equation is solved either by the FRW geometry, $h = 0$, or by an exponentially decreasing anisotropy, $h \propto {\rm e}^{-3 H t}$ (we disregard the slow roll decrease of $H$). 
This is at the basis of the cosmic  no-hair conjecture, according to which inflation is expected to rapidly erase any background anisotropy. 

 In the present model, with the scale factors appearing in $B^{ij}$ (see eq. (\ref{Bij})), also $\sigma$, and not only its derivative, is physical. To solve eq. (\ref{eq-sigma}), we perform the ansatz
\begin{equation}
\sigma \left( t \right) \propto  {\rm e}^{\int^t d t' \lambda \left( t' \right) H \left( t' \right)  }  
\;\;\Rightarrow\;\; \frac{\dot{\lambda}}{H} + \lambda^2 + \left( 3 - \epsilon \right) \lambda + 4 \epsilon c_T^2 = 0 \,\,.
\end{equation}
where we recall that $\epsilon,H, c_T$ in this equation are evaluated on the FRW geometry. We solve this equation to leading order in the slow roll parameters.  We obtain   
\begin{equation}
\lambda_1 = -3 + c_1 \left( t \right)  \epsilon   + {\rm O } \left( \epsilon^2 \right)           \;\;,\;\;
\lambda_2 = c_2 \left( t \right) \epsilon   + {\rm O } \left( \epsilon^2 \right) \,\,,
\end{equation}
where, in turns,   
\begin{equation}
\frac{\dot{c_1}}{H} + 6 - 3 c_1 + 3 c_L^2 = 0  \;\;,\;\;
\frac{\dot{c_2}}{H} + 3 c_2 + 4 c_T^2 = 0 \,\,.
\end{equation}
 Given eqs. (\ref{epsilon-eta}) and  (\ref{cTcL}), and given  that $Y$ and $Z$ are constant on a FRW background,  it is very reasonable to assume that   $\dot{c}_{L,T} = {\rm O } \left( \epsilon H  \right)$ or less. In this case, ~\footnote{We stress that the prolonged anisotropy is not consequence of  $\dot{c}_T = {\rm O } \left( \epsilon H  \right)$. 
 Even if  $\dot{c}_T = {\rm O } \left(  H  \right)$, the exponent $\lambda_2 = {\rm O } \left( \epsilon \right)$, which guarantees a slow isotropization.}   
\begin{equation}
c_1 = 2 + c_L^2 \;\;,\;\; c_2 = - \frac{4}{3}  c_T^2 \,\,.
\end{equation}

Therefore, to leading order, the anisotropy evolves as  
\begin{equation}
\sigma \left( t \right) \simeq \sigma_1 {\rm e}^{- \int \left[ 3 - \left( 2 + c_L^2 \right) \epsilon  \right]  H d t} + \sigma_2  {\rm e}^{-\int \frac{4}{3} c_T^2 \epsilon  H d t} \;\;,\;\; \sigma,\epsilon \ll 1 \,\,,
\label{bck-sol}
\end{equation}
where $\sigma_1$ and $\sigma_2$ are integration constant. The first term is (up to the subleading slow-roll correction) the fast decreasing solution found in standard slow roll inflation. The second term is a new, slowly decreasing solution, that is peculiar of this model.  We note that, as we anticipated, the coefficient $\lambda_2$ is proportional to the sound speed of the transverse modes; this testifies that the evolution of the phonons and of the anisotropy are determined by how the medium reacts to deformations. As we already showed, the solid that drives inflation needs to be extremely inefficient in responding to changes in the volume, and to the anisotropy.

\begin{widetext}
\begin{figure*}[ht!]
\centerline{
\includegraphics[width=0.6\textwidth,angle=0]{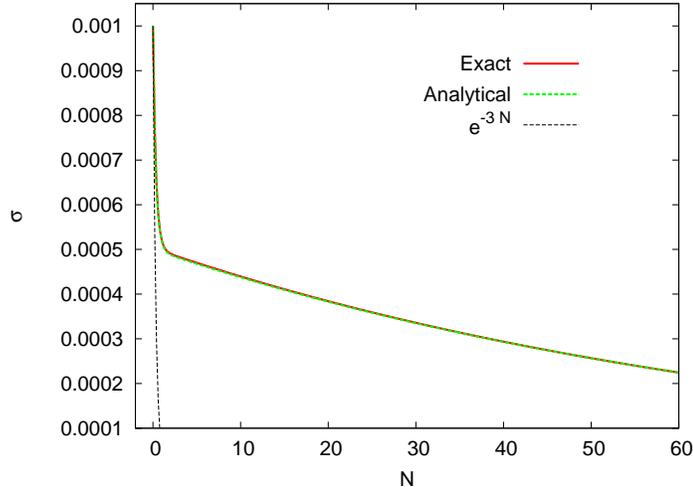}
}
\caption{Evolution of the anisotropy $\sigma$ (defined in eq. (\ref{bianchi})) as a function of the number of e-folds (of the average scale factor $\alpha$)  for the model (\ref{example-F}), and for  an initial approximately equal admixture of the two modes in (\ref{bck-sol}). The analytical solution (\ref{bck-sol}) shows a perfect agreement with the exact one. The line $\propto {\rm e}^{-3 N}$ shows the decrease of the fast decreasing mode. We note that this is also the rate at which the anisotropy  $\dot{\sigma}$ decreases    in standard inflationary  models.}
\label{fig:bck}\end{figure*}
\end{widetext}

In Figure \ref{fig:bck} we compare the approximate solution (\ref{bck-sol}) for the anisotropy against the exact solution obtained by numerically evolving the system  (\ref{ein-bianchi}). We choose the simplest possibility
\begin{equation}
F = F_0 X^\epsilon \,\,.
\label{example-F}
\end{equation}
 It is immediate to verify that,  at ${\rm O } \left( \sigma^0 \right)$, the parameter $\epsilon$ introduced in this function coincides  with the slow roll parameter $\epsilon = - \frac{\dot{H}}{H^2}$. With this choice, we obtain the sound speeds $c_T^2 = 1$, and $c_L^2 = \frac{1}{3}$.

In the evolution shown in the Figure, we choose $\epsilon =   0.01$ and $\sigma_{\rm in} = 0.001$. We want to verify the validity of the approximate analytical solution (\ref{bck-sol}). Therefore, we assume that it is valid, and we choose an initial condition that is an approximately equal admixture of both modes in  (\ref{bck-sol}). We do so by taking the initial value  $\dot{\sigma}_{\rm in} = \frac{\sigma_{\rm in}}{2} H \left[ - 3 + \left( 2 + c_L^2 \right) \epsilon - \frac{4}{3} c_T^2 \epsilon \right]$. We then use the first equation in (\ref{ein-bianchi}) to set the initial condition $\dot{\alpha}_{\rm in}$, and evolve numerically the last two equations in  (\ref{ein-bianchi}) (for the choice 
(\ref{example-F}), the initial value $\alpha_{\rm in}$ can be reabsorbed in $F_0$, which can then be rescaled away from the system of equations). If eq. (\ref{bck-sol}) is a good approximation of the exact solution, we must obtain that the  anisotropy  drops to about half its initial value within the first few e-folds~\footnote{The number of e-folds shown in the figure is $N = {\rm e}^{\alpha-\alpha_{\rm in}}$. It is accurate to use the ``average'' expansion rate as a measure of the expansion, since the anisotropy is extremely small,  $\dot{\sigma} \ll \dot{\alpha}$.} - corresponding to the fast decreasing component in (\ref{bck-sol}) -  followed by a much smaller decrease - corresponding to the second term in (\ref{bck-sol}). This is precisely what  the evolution in the figure shows. More precisely, we show both the exact numerical solution, and the analytic solution  (\ref{bck-sol}), and we see that the analytic solution is in excellent agreement with the exact one. In the figure, we also show the curve $\sigma = \sigma_{\rm in} {\rm e}^{-3 N}$. This curve has the same decrease of the fast decreasing mode. As we discussed, this reproduces the decrease of the anisotropy in standard models of scalar field inflation.  

Finally, we note that, strictly speaking, inflation never terminates for the choice (\ref{example-F}). As in ref. \cite{Endlich:2012pz}, we are assuming that  (\ref{example-F}) describes the function only for a  finite range of $X$, and then inflation terminates due to a change of $F$ (for example, due to a phase transition that transforms the solid into a fluid  \cite{Endlich:2012pz}). In the  evolution shown, we are simply following the evolution of the anisotropy for $60$ e-folds of inflation. For a longer duration of inflation, one finds that  the fast decreasing mode of (\ref{bck-sol}) has already decreased to negligible values during the entire last $\sim 60$ e-folds of inflation.

\subsection{Comparison with Wald's isotropization theorem}

Ref. \cite{Wald:1983ky} showed that a Bianchi geometry (with the possible exception of the type-IX case) undergoes a rapid isotropization under the influence of a cosmological constant  plus a source that satisfies the dominant and strong energy conditions. It is instructive to understand how the theorem precisely works and  why it  does not apply to the present context. To do this, we first summarize the computation of 
 \cite{Wald:1983ky}, and we then discuss the specific case of anisotropic solid inflation. 
 
 In the case analyzed by  \cite{Wald:1983ky}, the energy momentum tensor acquires the form 
\begin{eqnarray}
T_{\mu \nu} & = &  - \Lambda M_p g_{\mu \nu} +  T_{\mu \nu}^{\rm 2nd \; source}  \,\,,   
\label{Tmunu-wald}
\end{eqnarray}
where the first term is the cosmological constant contribution, and  the second term satisfies the dominant and strong energy conditions   ${\cal D} \geq 0$ and ${\cal S } \geq 0 $, where 
\begin{eqnarray}
{\cal D} & \equiv & t^\mu \, t^\nu T_{\mu \nu}^{\rm 2nd \; source}  \;\;,  \nonumber\\
{\cal S} & \equiv & t^\mu t^\nu \left( T_{\mu \nu}^{\rm 2nd \; source} - \frac{T^{\rm 2nd \; source} }{2} g_{\mu \nu} \right)  \,\,,
\label{D-S-def}
\end{eqnarray}
and where  $t^\mu$ is any  time-like future-directed vector. 

Ref. \cite{Wald:1983ky} contracted the Einstein equations with a normal vector $n^\mu$, to obtain their equations  (9) and (10).
In the Bianchi-I geometry (\ref{bianchi}) and in our notation, these equations read, respectively, 
\begin{eqnarray}
&& K^2 - 3 \Lambda - \frac{3}{2} \sigma^{\mu \nu} \sigma_{\mu \nu}  - \frac{3 \,    {\cal D}  }{M_p^2} = 0 \,\,, \nonumber\\
& &  \frac{d}{d t} K - \Lambda  + \frac{K^2}{3} +  \sigma^{\mu \nu} \sigma_{\mu \nu} +    \frac{ {\cal S}}{M_P^2} = 0 \,\,, 
\label{Wald-9-10}
\end{eqnarray}
where we have set $n^\mu = t^\mu =    \left\{ 1 , 0 , 0 , 0 \right\}$. In this expression, $K$ and $\sigma_{\mu \nu}$ are, respectively, the trace and the trace-free part of the   extrinsic curvature  on surfaces orthogonal to $n^\mu$. For  us , $K = 3 \dot{\alpha}$, and $\sigma^{\mu \nu} \sigma_{\mu \nu} = 6 \dot{\sigma}^2$, which are, respectively, the isotropic and anisotropic Hubble rates in (\ref{bianchi}). As long as the dominant and strong energy condition hold, ${\cal D} , {\cal S} \geq 0$, the two equations (\ref{Wald-9-10}) imply  \cite{Wald:1983ky} 
\begin{equation}
{\cal D} , {\cal S} \geq 0 \;\;\Rightarrow\;\; K > \sqrt{3 \Lambda} \;\;,\;\; \frac{1}{K^2 - 3 \Lambda} \, \frac{d K}{d t } \leq - \frac{1}{3} \,\,.
\label{wald-here}
\end{equation}
The second relation can be then integrated and combined with the first one  to show that $K \rightarrow \sqrt{3 \Lambda}$ with exponential accuracy on a timescale $\sqrt{3/\Lambda}$    \cite{Wald:1983ky}. Inserting this result into the first of (\ref{Wald-9-10}), we then see that $\sigma^{\mu \nu} \sigma_{\mu \nu} \rightarrow 0$ on the same timescale. We thus recover an (isotropic) de Sitter expansion driven by $\Lambda$  \cite{Wald:1983ky}. The inequalities (\ref{wald-here}) play a crucial role for this result. We stress that they  are a consequence of the dominant and strong energy conditions.

Let us now discuss solid inflation, for which the  energy momentum     is given in eq.  (\ref{Tmunu}). Strictly speaking, this is not the energy momentum tensor of a cosmological constant plus a second source; however, given that the model supports inflation, it  still proves useful for the comparison  with \cite{Wald:1983ky}  to use this two component decomposition as an effective description. The form of   (\ref{Tmunu}) would  suggest to identify the first term as the cosmological constant contribution. However, the function $F$ is not constant (but rather slow roll evolving), and the proof in  \cite{Wald:1983ky} would not apply. We therefore decompose eq. (\ref{Tmunu}) as
\begin{eqnarray}
T_{\mu \nu} & = & g_{\mu \nu} F \left( t_0 \right) + \left\{  g_{\mu \nu} \left[  F \left( t \right) -  F \left( t_0 \right) \right]
 - 2 \partial_\mu \phi^i \partial_\nu \phi^j \, \frac{\partial F}{\partial B^{ij}} \right\}  \nonumber\\
& \equiv & - \Lambda M_p^2 g_{\mu \nu} +  T_{\mu \nu}^{\rm 2nd \; source}  \,\,, 
\label{Tmunu2}
\end{eqnarray}
where $\Lambda \equiv - F \left( t_0 \right) / M_p^2 > 0$, and where $t_0$ is a fixed time during inflation, say the starting time, at which the geometry is of the Bianchi-I type, and one is interested in whether the rapid isotropization takes place.

 Inserting (\ref{Tmunu2}) into (\ref{D-S-def}) we obtain
\begin{eqnarray}
{\cal D} & = & - F \left( t \right) + F \left( t_0 \right) \,\,, \nonumber\\
{\cal S} & = &  F \left( t \right) - F \left( t_0 \right) - F_X \left( \frac{1}{a^2} + \frac{2}{b^2} \right) \;\;,
\end{eqnarray}
and, using these expressions, we can readily verify the  system (\ref{Wald-9-10}) is equivalent to the three  background equations (\ref{ein-bianchi}) (we recall that only two of these equations are independent). We have already solved these equations   in the first part of this Section, and we have  obtained that the anisotropy is not erased on the timescale  $\sqrt{3/\Lambda}$. The technical reason 
for this is that  ${\cal D} < 0$ in this model (while instead ${\cal S} > 0$). This is due to the fact that $F$ is negative and it decreases in magnitude during inflation. As a consequence, the two conditions (\ref{wald-here}) do not hold.

We have therefore shown that the total energy momentum tensor of solid inflation cannot be rewritten as the sum of a cosmological constant plus a second term that satisfies the dominant and strong energy conditions, which explains why Wald's theorem does not apply.  One may worry that the failure of the dominant energy  condition might be a signal of instability. This is not the case, since the split in (\ref{Tmunu2}) is only an effective description to be able to compare with the premise of Wald's theorem, but there is no instability associated with the full energy momentum tensor.

\section{Scalar curvature perturbations on the anisotropic solution }
 \label{sec:Bianchi-perts}

We now compute the   primordial perturbation ${\hat \zeta}$ on the anisotropic background  obtained in the previous Section.
 As we shall see, the observed statistical isotropy of the CMB perturbations forces the background anisotropy to be small, $\sigma \ll 1$.
Therefore, we can compute ${\hat \zeta}$ in a perturbative expansion around the FRW solution studied in Section \ref{sec:FRW}. \footnote{In this Section, ${\hat \zeta}$ (respectively  ${\hat \zeta}^{(0)}$) denotes the curvature perturbation of the anisotropic background (resp. on the FRW background). The hat denotes the quantum operator for the curvature, expanded in terms of annihilation / creation operators and of the mode function  $\zeta$ (resp. $\zeta^{(0)}$), see eq. (\ref{zeta-deco}). The FRW mode function 
 $\zeta^{(0)}$ is given in   (\ref{explicit-zetasol}), where it was denoted without the $(0)$ suffix.} We perform the computation  through the in-in formalism:
\begin{eqnarray}
&& \left\langle {\hat \zeta}_{\vec{k}_1} \,  {\hat \zeta}_{\vec{k}_2}  \left( \tau \right)  \right\rangle = \sum_{N=0}^\infty \left( - i \right)^N \int^\tau  d \tau_1 \dots   \int^{\tau_{N-1}}  d \tau_N \nonumber\\
&& \;\;\;  \left\langle \left[ \left[ \dots  \left[ {\hat \zeta}_{\vec{k}_1}^{(0)} \,  {\hat \zeta}_{\vec{k}_2}^{(0)}  \left( \tau \right) ,\;
H_{\rm int} \left( \tau_1 \right) \right] , \dots \right] ,\, H_{\rm int} \left( \tau_N \right) \right] \right\rangle \nonumber\\
\label{in-in}
 \end{eqnarray} 
where $H_{\rm int} = - \int d^3 x {\cal L}_{\rm int}$, and ${\cal L}_{\rm int}$  is the quadratic lagrangian for the perturbations on the Bianchi background minus the quadratic lagrangian on a FRW background (we disregard terms that are higher order than quadratic in the perturbations inside ${\cal L}_{\rm int}$). We note that each term in  ${\cal L}_{\rm int}$ can be written as an expansion series in the anisotropy $\sigma$, that, in general, starts at ${\rm O } \left( \sigma \right)$.

\begin{widetext}
\begin{figure*}[ht!]
\centerline{
\includegraphics[width=\textwidth,angle=0]{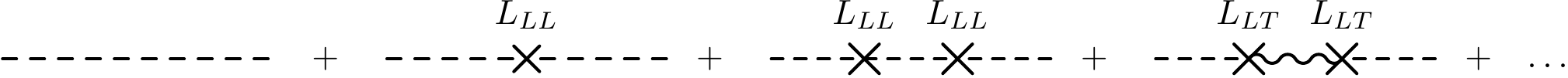}
}
\caption{Leading diagrams for $\langle {\hat  \zeta}^2 \rangle$ on an anisotropic background. 
The first diagram is the FRW result, while the second diagram is the linear correction in the anisotropy.
Only these two diagrams are computed in the main text. We disregard quadratic (the last two diagrams shown) and higher order corrections in the anisotropy.
}
\label{fig:zz}\end{figure*}
\end{widetext}

In the in-in formalism, perturbations are quantized in the interaction picture: this means that, in our computation, 
the FRW quantization of  \cite{Endlich:2012pz} applies.  However, due to the anisotropy, the scalar/vector/tensor   perturbations are no longer decoupled from each other in the full quadratic action, and this gives rise to additional terms in  ${\cal L}_{\rm int}$. Due to the residual SO(2) background  isotropy of (\ref{bianchi}), one mode of $\pi_T^i$ and one mode of $h_{ij}$ remain decoupled from $\pi_L$ at the quadratic level
\cite{Gumrukcuoglu:2007bx}. Therefore, ${\cal L}_{\rm int}$ couples $\zeta^{(0)}$ with one mode of $\pi_T^i$ and one mode of $h_{ij}$. 
Since ${\cal L}_{\rm int}$ is quadratic in the fields, its terms can be diagrammatically visualized as the ``mass insertions'' $L_{LL}$ (terms involving two scalar modes),  $L_{LT}$ (terms involving one scalar and one vector mode),   $L_{LH}$ (terms involving one scalar and one tensor mode), $L_{TT}$, $L_{TH}$, and $L_{HH}$. Figure (\ref{fig:zz}) shows some of the leading order contributions to $\langle {\hat \zeta}^2 \rangle$ arising when these mass insertions are used in (\ref{in-in}) (the variable $N$ in  (\ref{in-in}) coincides with the number of mass insertions present in the diagram). In the Figure, dashed lines denote the scalar mode; curved line denotes the vector mode, and the crosses denote  mass insertions.

It is clear from the Figure that the interactions between the scalar mode ($L$) and one of the other two modes ($T$ or $H$) contribute to $\langle {\hat \zeta}^2 \rangle$ only at ${\rm O } \left( \sigma^2 \right)$ or higher. Therefore, if  $L_{LL}$ provides the only   ${\rm O } \left( \sigma \right)$ contribution to $\langle {\hat \zeta}^2 \rangle$, it is the dominant correction to the power spectrum of ${\hat \zeta}$ due to the anisotropy.  We now  compute this contribution. We do so in two Subsections. In Subsection \ref{subs:Hint} we compute the interaction hamiltonian. In Subsection \ref{subs:dP2} we insert the interaction hamiltonian in  (\ref{in-in}) and evaluate the  correction of the power spectrum due to the anisotropy.

\subsection{Computation of $H_{\rm int}$}
\label{subs:Hint}

To obtain $L_{LL}$, we set to zero all the perturbations apart from the scalar one. For the three scalar fields, this means
\begin{equation}
\phi^i = x^i - 3 i  \int \frac{d^3 k}{\left( 2 \pi \right)^{3/2} } {\rm e}^{i \vec{k} \cdot \vec{x}} \frac{  k^i}{k^2} \, {\hat \zeta} \left( t , \vec{k} \right) \,\,,
\label{scalar-x}
\end{equation} 
where the relation (\ref{zeta}) has been used. 

We recall that we are working in the spatially flat gauge, so that the spatial part $g_{ij}$ of the metric  is given by (\ref{bianchi}). We instead introduce perturbations in $g_{00} = - 1 - 2 \Phi$, and $g_{0i} = \delta g_{0i}   $, which need to be retained as they are nondynamical and are algebraically given in terms of ${\hat \zeta}$ and $\dot{\hat \zeta}$ (from the linearized Einstein equation, which is equivalent to extremizing the quadratic action of the perturbations with respect to them).  

We then evaluate the action up to second order in the perturbations, and integrate out the nondynamical modes in $\delta g_{0\mu}$. 
The solutions for $\Phi$ and $\delta g_{0i}$ in terms of ${\hat \zeta}$ and  $\dot{\hat \zeta}$ are rather lengthy and not illuminating, and so we do not explicitly report them here. We insert the solutions back in the quadratic action, which then becomes the action for the dynamical mode ${\hat \zeta}$ only. This is the standard procedure to obtain the quadratic action for the perturbations of any system.  The resulting expression is formally of the type
\begin{eqnarray}
S \left[ {\hat \zeta} \right] & = & 
\int d t d^3 k \Bigg\{ f_{\rm kin} \left[ \alpha , \dot{\alpha} , \sigma , \dot{\sigma} \right]  \vert \dot{{\hat \zeta}} \vert^2 
 +   f_{\rm mas} \left[ \alpha , \dot{\alpha} , \sigma , \dot{\sigma} \right]  \vert {\hat \zeta} \vert^2   \nonumber\\
& & \quad\quad \quad\quad + 
\left( f_{\rm mix} \left[ \alpha , \dot{\alpha} , \sigma , \dot{\sigma} \right]  \dot{{\hat \zeta}}^* {\hat \zeta} + {\rm h. c. } \right)   \Bigg\} \,\,,
\label{formal-Szeta}
 \end{eqnarray} 
where the three functions are functions of the background (we eliminate $\ddot{\alpha}$ and $\ddot{\sigma}$  from these expressions by the use of the background equations of motion (\ref{ein-bianchi}); specifically, we enforce the background equations by expressing $\ddot{\alpha}, \ddot{\sigma}$ and $F$ as a function of the other quantities. Thanks to this, we are sure that our expressions cannot be further simplified by the use of the background equations). 
 
 The explicit expressions for these three functions (that we obtained by the use of Mathematica), are extremely lengthy, and not illuminating, and for this reason we do not report them here.  We expand these expressions in the anisotropy parameter $\sigma$, and obtain an expansion of the action $S \left[ {\hat  \zeta } \right]$ in the anisotropy.
 We formally  write the resulting 
expression  as
\begin{equation}
S \left[ {\hat \zeta} \right] = \sum_{n=0}^\infty  S^{(n)} \left[ {\hat \zeta} \right] \,\,,
\end{equation}
where  $ S^{(n)} $ is of order $n$ in the anisotropy. Namely, it is obtained by the    ${\rm O } \left( \sigma^n , \sigma^{n-1} \dot{\sigma} ,   \sigma^{n-2} \dot{\sigma}^2 ,      \dots , \dot{\sigma}^n \right)$ expressions for $f_{\rm kin}$,  $f_{\rm mix}$, and   $f_{\rm mas}$. We verified that, as it must be,  the zeroth-order action $S^{(0)}$ coincides with (\ref{azpi-FRW}). According to the above discussion, we are only interested in the explicit expression for  $S^{(1)}$, as this is the term that gives $L_{LL}$ at first order in the anisotropy. Inside  $S^{(0)}+S^{(1)}$,  the following functional  derivatives of $F$ appear: $F_X, F_Y, F_Z, F_{XX}, F_{XY}, F_{XZ}$. These functions can be evaluated in the   FRW background, as the three invariants $X,Y,Z$ in the Bianchi geometry coincides with that in the  FRW geometry up to ${\rm O } \left( \sigma^2 \right)$.  We eliminate the mix term $\propto f_{\rm mix}$ from this expression through an integration by parts. This introduces the three derivatives $\frac{d}{d t} F_X,  \frac{d}{d t} F_Y$, and $ \frac{d}{d t} F_Z$, which we evaluate through  $\frac{d}{d t} F_i = \dot{X} \, F_{iX} \cong - 6 {\rm e}^{-2 \alpha} \dot{\alpha }  \, F_{iX} $ (where 
$\cong$ indicates that the two expression coincide up to second order corrections in the anisotropy). 

Therefore, proceeding as just indicated, we obtain an expression for  $S^{(1)}$ where $F$ only explicitly enters through its   $F_X, F_Y, F_Z, F_{XX}, F_{XY}, F_{XZ}$ derivatives, evaluated on the FRW background. It is useful to rewrite these derivatives in terms of more immediate physical parameters, as the slow roll parameters and the sound speed. Using (\ref{epsilon-eta}) and (\ref{cTcL}), we can write
\begin{eqnarray}
F_X & = & - {\rm e}^{2 \alpha} M_p^2 \epsilon \dot{\alpha}^2 \,\,,  \nonumber\\
F_{XX} & = &  \frac{1}{6} {\rm e}^{4 \alpha} M_p^2 \epsilon \left(  2 - 2 \epsilon + \eta \right) \dot{\alpha}^2 \,\,, \nonumber\\
F_Y + F_Z & = &  \frac{9}{2} M_p^2 \left(  1 - c_T^2   \right) \epsilon \dot{\alpha}^2 \,\, .
\label{Fexp1}
\end{eqnarray}
The combination $F_Y-F_Z$ is not related to any background quantity defined above. In analogy with the last of (\ref{Fexp1}) we define
\begin{eqnarray}
F_Y - F_Z & \equiv &  \frac{9}{2} M_p^2 \mu  \epsilon \dot{\alpha}^2 \,\, .
\label{Fexp2}
\end{eqnarray}
where it is reasonable to assume that also $\mu$ is of order one, and slowly varying. Differentiating the last of (\ref{Fexp1}) we obtain
\begin{equation}
F_{XY} + F_{XZ} = \frac{3}{4} {\rm e}^{2 \alpha} M_P^2 \epsilon \left[ 2 \epsilon - \eta + c_T^2 \left( 2 s_T - 2 \epsilon + \eta \right) \right] \dot{\alpha}^2 \,\,, 
\label{Fexp3}
\end{equation}
where, in analogy to \cite{Endlich:2012pz}, we have defined the slow roll quantity $s_T \equiv \frac{\dot{c}_T}{\dot{\alpha} c_T} \,$.
Finally, we find that $F_{XY} - F_{XZ} $ does not enter in  $S^{(1)}$.

Using these expressions, the  ${\rm O } \left( \sigma \right)$ action for ${\hat \zeta}$ acquires the form (\ref{formal-Szeta}), with
\begin{eqnarray}
f_{\rm kin}^{(1)} & = & 18 {\rm e}^{3 \alpha} M_p^2 \epsilon \dot{\alpha}^2 P_2 \left( {\rm cos } \, \theta \right)
\frac{2 p^2 \left( c_T^2-1 \right) \sigma + 3 \epsilon \dot{\alpha} \dot{\sigma} }{\left( p^2 + 3 \epsilon \dot{\alpha}^2 \right)^2} \,\,, \nonumber\\ 
f_{\rm mas}^{(1)} & = &  \frac{3 {\rm e}^{3 \alpha} M_p^2 \epsilon \dot{\alpha}^2}{\left( p^2 + 3 \epsilon \dot{\alpha}^2 \right)^3}  P_2 \left( {\rm cos } \, \theta \right) \left[ c_6 p^6 + c_4 p^4 + c_2 p^2 + c_0 \right] \,\,,  \nonumber\\
\label{S-zeta1}
\end{eqnarray}
with
\begin{eqnarray}
c_6 & = & 4  \left[  5 - 2 c_T^2\left(2-2 s_T+2 \epsilon-\eta\right) - \mu \right] \sigma \,\,, \nonumber\\
c_4 & = & 12 \epsilon \left[  20 - 4 \epsilon + 2 \eta - 2 c_T^2 \left( 11 - 4 s_T + 2 \epsilon - \eta \right) - 3 \mu \right] \dot{\alpha}^2  \sigma \nonumber\\
& & + 6 \epsilon \left[  3 - 4 c_T^2 - 2 \epsilon + \eta \right] \dot{\alpha} \dot{\sigma} \,\,, \nonumber\\
c_2 & = & 36 \epsilon^2 \left[ 16 -  c_T^2 \left( 21 - 6 s_T + 4 \epsilon - 3 \eta \right) - 3 \mu \right] \dot{\alpha}^4 \sigma \nonumber\\
& & - 9 \epsilon^2 \left( 2 + 12 c_T^2 - 4 \epsilon + 3 \eta \right) \dot{\alpha}^3 \dot{\sigma} \,\,, \nonumber\\ 
c_0 & = & 108 \epsilon^3 \left[ 5 - c_T^2 \left( 7 - 2 s_T - \eta \right) - \mu \right] \dot{\alpha}^6 \sigma \nonumber\\
& & - 27 \epsilon^3 \left( 4 c_T^2 + \eta \right) \dot{\alpha}^5 \dot{\sigma} \,\,.
\end{eqnarray}
In (\ref{S-zeta1}),   $p \equiv k {\rm e}^{- \alpha}$ is the physical momentum of the mode, $\theta$ the angle between the direction of the momentum and the anisotropic direction ${\hat x}$, and $P_2$ is the  Legendre polynomial of order two. Moreover, we recall that $f_{\rm mix} = 0$ thanks to the integration by parts.

As we are interested in the perturbations around the slowly decreasing  anisotropic solution obtained in the previous Section, we set 
$\dot{\sigma} \cong - \frac{4}{3} c_T^2 \epsilon H \sigma$ (we recall that $H = \dot{\alpha}$).  The expressions (\ref{S-zeta1}) then become
\begin{eqnarray}
f_{\rm kin}^{(1)} & \cong & - 36 {\rm e}^{3 \alpha} M_p^2 \epsilon H^2 P_2 \left( {\rm cos } \, \theta \right) \sigma \frac{\left( 1 - c_T^2 \right) p^2 + 2 c_T^2 \epsilon^2 H^2}{\left( p^2 + 3 \epsilon H^2 \right)^2} \,\,,  \nonumber\\
f_{\rm mas}^{(1)} & \cong &  \frac{12 {\rm e}^{3 \alpha} M_p^2 \epsilon H^2 \sigma}{\left( p^2 + 3 \epsilon H^2 \right)^3}  P_2 \left( {\rm cos } \, \theta \right) \Bigg[   p^6 \left( 5 - 4 c_T^2 - \mu \right) \nonumber\\
& & \!\!\!\!\!\!\!\! \!\!\!\! + 3  \epsilon H^2 p^4  \left( 20 - 22 c_T^2 - 3 \mu \right) 
+ 9  \epsilon ^2 H^4 p^2  \left( 16 - 21 c_T^2 - 3 \mu \right) \nonumber\\
& & \!\!\!\!\!\!\!\! \!\!\!\!  + 27 \epsilon^3 H^6  \left( 5 - 7 c_T^2 - \mu \right) \Bigg] \,\,,
\end{eqnarray}
where we have disregarded terms of ${\rm O } \left( \epsilon, \eta, s_T \right)$ or higher when compared with ${\rm O } \left( 1 \right)$ terms.

Finally, switching to conformal time, from the expression (\ref{formal-Szeta}) we obtain the interaction  hamiltonian
\begin{eqnarray}
H_{\rm int}^{(1)} \left( \tau \right) & = & - \int d^3 k \Bigg[ {\rm e}^{- \alpha} f_{\rm kin}^{(1)} {\hat \zeta}^{(0)'}_{-\vec{k}} \left( \tau    \right)  {\hat \zeta}^{(0)'}_{\vec{k}} \left( \tau   \right) 
\nonumber\\
& & + {\rm e}^\alpha  f_{\rm mas}^{(1)} {\hat \zeta}^{(0)}_{-\vec{k}} \left( \tau    \right)  {\hat \zeta}^{(0)}_{\vec{k}} \left( \tau   \right) \Bigg] + {\rm O } \left( \sigma^2 \right) \,\,.
\label{Hint-formal}
\end{eqnarray} 

\subsection{Evaluation of the  power spectrum}
 \label{subs:dP2}

We now insert (\ref{Hint-formal}) into  (\ref{in-in})  , to obtain
\begin{eqnarray}
& & \left\langle {\hat \zeta}_{\vec{k}_1} \,  {\hat \zeta}_{\vec{k}_2}  \left( \tau \right)  \right\rangle = 
 \left\langle {\hat \zeta}_{\vec{k}_1}^{(0)} \,  {\hat \zeta}_{\vec{k}_2}^{(0)}  \left( \tau \right)  \right\rangle      \nonumber\\
 & & \quad \quad 
 - i \int^\tau d \tau_1 
\left\langle \left[  {\hat \zeta}_{\vec{k}_1}^{(0)} \,  {\hat \zeta}_{\vec{k}_2}^{(0)}  \left( \tau \right)  , H_{\rm int}^{(1)} \left( \tau_1 \right) 
\right] \right\rangle + {\rm O } \left( \sigma^2 \right) \,\,. \nonumber\\
\label{inin01}
\end{eqnarray}

To evaluate this expression, we decompose the quantum field ${\hat \zeta}_{\vec{k}}$ into
\begin{equation}
{\hat \zeta}_{\vec{k}} \left( \tau \right)  = \zeta_{\vec{k}} \left( \tau \right) a_{\vec{k}} +  \zeta_{-\vec{k}}^* \left( \tau \right) a_{-\vec{k}}^\dagger
\;\;,\;\; \left[ a_{\vec{k}} , a_{\vec{k}'} \right] = \delta^{(3)} \left( \vec{k} + \vec{k'} \right) \,\,,
\label{zeta-deco}
\end{equation}
and identically for ${\hat \zeta}^{(0)}$. 

The two point correlation function is related to the power spectrum by
\begin{eqnarray}
\langle {\hat \zeta}_{\vec{k}_1} \left( \tau \right)  {\hat \zeta}_{\vec{k}_2} \left( \tau \right) \rangle \equiv 2 \pi^2 \frac{\delta^{(3)} \left( \vec{k}_1 + \vec{k}_2 \right)}{k_1^3 } \, P_\zeta \left( \vec{k}_1 \right) \;\;,
\end{eqnarray}
and we finally define
\begin{equation}
 P_\zeta \left( \vec{k} \right)= P_\zeta^{(0)} \left( k \right) +  P_\zeta^{(1)} \left( \vec{k} \right) + {\rm O } \left( \sigma^2 \right) \,\,,
 \end{equation}
corresponding, respectively, to the unperturbed FRW correlator and to the first order correction in $\sigma$.  To leading order in slow roll,  the FRW expression (\ref{zetalarge-gen}) gives
\begin{equation}
P_\zeta^{(0)} \left( \vec{k} \right) = \frac{k^3}{2 \pi^2} \left\vert \zeta_k^{(0)} \right\vert^2 \simeq \frac{ 1}{8 \pi^2 c_L^5} \, \frac{H^2}{M_p^2  \epsilon } \;\;,\;\; - c_L  k \tau \ll 1 \,\,.
\label{P0}
\end{equation}

For the first order correction, evaluating the commutator and the expectation value in (\ref{inin01}) we obtain
\begin{eqnarray}
P^{(1)}_\zeta \left( \vec{k} \right) & = & \frac{ 2 k^3}{\pi^2} \; \times \nonumber\\
& & \!\!\!\!\!\!\!\! \!\!\!\!  \!\!\!\!  \!\!\!\!  \!\!\!\!  {\rm Im } \left[  \zeta^{(0)* 2}_k \left( \tau \right) \int^\tau d \tau_1
\left( {\rm e}^{-\alpha} \, f_{\rm kin}^{(1)}  \zeta^{(0)'2} + {\rm e}^\alpha f_{\rm mas}^{(1)}   \zeta^{(0)2} \right)_{\tau_1,\vec{k}} \right] \,\,. \nonumber\\
\end{eqnarray}

We inserted the solution (\ref{explicit-zetasol})-(\ref{norma-zeta}) into this expression. We could not perform the time integration in an exact closed form, and we therefore divided the integral into the two regimes $-\infty < \tau_1 < - {\rm O } \left( \frac{1}{c_L \, k} \right) $, and $ - {\rm O } \left( \frac{1}{c_L k} \right) < \tau_1 < \tau$. In the first regime, we used the sub-horizon limit of  (\ref{explicit-zetasol}) for $\zeta \left( \tau_1 \right)$ and its derivative, while in the second regime we used the super-horizon limit. For $\zeta \left( \tau \right)$ we instead use the super horizon limit of  (\ref{explicit-zetasol}), given that we are interested in the super-horizon value for $P^{(1)}_\zeta$. Proceeding in this way, we obtain the estimate
\begin{equation}
\frac{ P^{(1)}_\zeta \left( \vec{k} \right) }{ P^{(0)}_\zeta \left( k \right) } = P_2 \left( \cos \, \theta \right) \sigma \left[ \left( 1 + 24   c_L^2 + 4 \mu^2 \right) \epsilon N_{\rm cmb} + {\rm O } \left( 1 \right) \right]
\label{estimate-P1}
\end{equation}
where the first term in the square parenthesis is the contribution from the late time integration limit $\tau_1 \la \tau$, while the second term is the contribution from $\tau_1$ in the sub-horizon regime and at horizon crossing. The second contribution may be the dominant one, so we regard (\ref{estimate-P1}) as an estimate of  $P^{(1)}_\zeta$, which we will use to set an order of magnitude upper bound on the anisotropy parameter $\sigma$.    We recall that $P_2$ is the Legendre polynomial of order two, while $\theta$ is the angle between $\vec{k}$ and the anisotropic direction. The quantity $\sigma$, as well as the other quantities on the right hand side of (\ref{estimate-P1}) are the values assumed at horizon crossing. 

 Therefore our estimate for the power spectrum of $\zeta$ on super-horizon scales is
\begin{equation}
\label{Ptot}
P_\zeta \left( \vec{k} \right) =   \frac{ 1}{8 \pi^2 c_L^5} \, \frac{H^2}{M_p^2  \epsilon } \left[ 1 + {\rm O } \left( 1 \right) \sigma P_2 \left( \cos \theta \right)  \right]\, .
\end{equation}
Let us conclude this section with a few comments. First, from Eq.~(\ref{Ptot}) we can read the anisotropic amplitude of the power-spectrum $g_*$ in the parameterization~\cite{Ackerman:2007nb}
\begin{equation}
P_\zeta \left( \vec{k} \right) = P \left( k \right) \left[ 1 + g_* \, \cos^2 \theta  \right]\, .
\label{acw}
\end{equation}
We find, in the phenomenologically allowed region $\vert g_* \vert \ll 1$, that  $g_* = O(1)\, \sigma$. Different limits have been obtained 
on such a parameter, starting from the analysis of the WMAP7  data  \cite{Komatsu:2010fb}
that gives $g_* = 0.29 \pm 0.031$ \cite{Groeneboom:2009cb}. Such a large effect has been clearly demonstrated to be due to beam asymmetries in WMAP9 data~\cite{Hanson:2009gu,Hanson:2010gu,Bennett:2012fp} and is not present in the Planck data \cite{Ade:2013nlj}. On different scales (and marginalizing over the preferred direction) Large-Scale Structure data analysis constrain $-0.41 < g_* < 0.38$ at $95 \%$ C.L. \cite{Pullen:2010zy} (the amplitude of the anisotropy may in general  be scale dependent~\cite{Ackerman:2007nb}). Therefore a $10\%$ level anisotropy, $|g_*|=0.1$ ($(1\%)$ level, $|g|_*=0.01$) would correspond to an anisotropy parameter $\sigma \simeq 0.1$ $(0.01)$. 

As a second comment, notice that $g_*$, being determined by $\sigma$, is not simply proportional to the 
``anisotropic Hubble rate'' $\Delta H/H = \dot{\sigma}/H$, as one might naively expect. Rather, since  $ \dot{\sigma} \propto c_T^2 \epsilon H \sigma$ (see eq.~(\ref{bck-sol})), $g_*$ turns out to be 
\begin{equation}
g_* = {\rm O } \left( \frac{\Delta H}{\epsilon \, H} \right) \gg 
 {\rm O } \left( \frac{\Delta H}{  H} \right) 
\label{gstar}
\end{equation}
 This is analogous to what happens in the $f\left( \phi \right) F^2$ models~\cite{Dulaney:2010sq,Gumrukcuoglu:2010yc,Watanabe:2010fh}.

As a third comment, we note that the final background anisotropy still present at the end of inflation may give rise to corrections to the variable $\zeta$ which are of ${\rm O } \left( \frac{\Delta H}{H} \vert_{\rm end} \right) = {\rm O } \left( \frac{\dot{\sigma}}{H}  \vert_{\rm end} \right) = {\rm O } \left( \epsilon_{\rm end} \sigma_{\rm end} \right)$, where the suffix ``end'' refers to the value assumed at the end of inflation. This, and - more in general - the dynamics of reheating after inflation, may generate corrections to the observed value of $P_\zeta$. 
  As discussed in  \cite{Endlich:2012pz}, 
it is reasonable to assume that in this model inflation is terminated by a phase transition, during which the solid decays into conventional matter. Ref.   \cite{Endlich:2012pz} computed the perturbations of solid inflation on an isotropic background, showing that $\zeta$ is continuous at this transition. Therefore, any correction to $g_*$ that emerges from these effects can be at most of ${\rm O } \left(  \sigma_{\rm end} \right)$ which is parametrically much smaller than the ${\rm O } \left( \frac{\sigma}{\epsilon} \right)$ value that we have studied and given in (\ref{gstar}).

Finally, we note that, while the  $f\left( \phi \right) F^2$ results in a negative $g_*$ \cite{Dulaney:2010sq,Gumrukcuoglu:2010yc,Watanabe:2010fh}, in our case both signs of $g_*$ are possible.

\section{Conclusions}
\label{sec:conclusions}

We showed that solid inflation supports prolonged anisotropic inflationary solutions. This constitutes a stable example  based on standard gravity and scalar fields only that violates the conditions of the  so called cosmic no-hair conjecture \cite{Wald:1983ky}. This result strengthens the analogy between solid inflation and the $f \left( \phi \right) F^2$ mechanism. It was already shown that both models exhibits a bispectrum with a nontrivial angular dependence in the squeezed limit. We have now shown that this analogy also holds at the background level, since  the $f \left( \phi \right) F^2$ mechanism also supports anisotropic inflation without instabilities.

In    this Section we discuss a few open questions on solid inflation. First of all, given the strong analogy between solid inflation and the  $f \left( \phi \right) F^2$ mechanism, both at the background level and at the level of the bispectrum, it would be interesting to explore whether the models have other similarities, and, in particular, whether they can be formulated within a unique effective description. For instance, ref.  \cite{Biagetti:2013qqa} showed how the previously obtained results for the $f \left( \phi \right) F^2$ models can be understood in terms of symmetries of the vector field. It may be possible that their computations can be further extended to include solid inflation as well,  perhaps  developing an effective field theory of broken spatial translational and rotational symmetries during inflation (analogously to the effective field theory that identifies the cosmological perturbations with the goldstone bosons of the broken time translational invariance in the standard cases \cite{Cheung:2007st,Weinberg:2008hq}). 

Possibly, the similarities between the two models will also include the infra-red sensitivity to anisotropic super-horizon modes that characterizes the  $f \left( \phi \right) F^2$ model \cite{Bartolo:2012sd}. Assume that inflation starts from an isotropic configuration, for instance  with a triad of orthogonal vectors of equal magnitude, and choose the function  $f \left( \phi \right)$ to produce a frozen scale invariant spectrum of vector perturbations  outside the horizon. Assume also that the total number of e-folds of inflation $N_{\rm tot}$ is  greater than the number of e-folds $N_{\rm CMB} \simeq 60$ at which the CMB modes left the horizon. 
 The modes of the vector fields that left the horizon in the first $\sim N_{\rm tot} - N_{\rm CMB}$ e-folds of inflation become classical at horizon exit and randomly add up with each other. 
This sum is not constant across the universe, but the nontrivial spatial-dependence takes place only on scales much greater than our current horizon, and therefore this nontrivial spatial dependence is unobservable. However, it is crucial to realize that the sum itself is  {\it not unobservable}. The modes that leave the horizon in the final $60$ e-folds see this sum as a classical homogeneous background quantity. This last statement is commonly accepted in the case of scalar fields (this is the origin of the coherent vev in the Affleck-Dine \cite{Affleck:1984fy} and in the curvaton \cite{Lyth:2001nq} mechanisms), but - as remarked in  \cite{Bartolo:2012sd} - its validity has nothing to do with the spin of the field, but only with the property of the super-horizon modes. Any field (of any spin) that has a frozen  spectrum of perturbations outside the horizon develops a coherent vev, that is locally observed as a homogeneous quantity. The only role played by the higher   spin is that, differently from a scalar field, a homogeneous vector breaks isotropy locally.  

The theory only provides a statistical prediction for this classical vector field $\vec{V}_{\rm IR}$: if we could observe many independent realizations of the first $N_{\rm tot} - N_{\rm CMB}$ e-folds of inflation, we would find a  (nearly) gaussian distribution for  $\vec{V}_{\rm IR}$ with zero mean and variance $\langle V_{\rm IR}^2 \rangle \propto N_{\rm tot} - N_{\rm CMB}$ \cite{Bartolo:2012sd}. However, we can observe only one realization, so we naturally expect to observe a vector with magnitude $\vert \vec{V}_{\rm obs} \vert \simeq  \sqrt{ \langle \vec{V}_{\rm IR}^2 \rangle } $. Even if classically one starts from an isotropic triad, there is no reason why the three infra-red sums of the different vectors should be equal to each other (each sums is the random addiction of quantum vectors, and no gauge symmetry can enforce that the quantum fluctuation of each mode of one vector is identical to the quantum fluctuation of each mode of another vector), and the natural statistical expectation for the difference is also given by $ \sqrt{ \langle \vec{V}_{\rm IR}^2 \rangle } $. This unavoidably generates an anisotropy for the classical vector background, which in turns imprints a strong anisotropy to the power spectrum of the inflation through its direct $f \left( \phi \right) F^2$ coupling to the vector. This results in a natural expectation for the duration of inflation in all models that support a scale invariant vector field outside the horizon, and, in particular, for all models of anisotropic inflation, anisotropic curvaton, and inflationary magnetogenesis \cite{Bartolo:2012sd}. In the $f \left( \phi \right) F^2 $ mechanism,  the anisotropy exceeds the $1\%$ level ($10\%$ level) if inflation lasted $\sim 5$ e-folds ($\sim 50$ e-folds) more than the minimal amount required to produce the CMB modes  \cite{Bartolo:2012sd}. 

It is possible that a similar problem also holds for solid inflation. This is not the anisotropy that we have studied in this work, as here we have assumed that the background is initially anisotropic, and we have followed the background evolution dictated by the classical equations of motion. However, there is no reason to expect that, even starting from an isotropic background, the three scalars of solid inflation will develop three identical power spectra. The difference will be encoded both in the longitudinal and in the vector modes of the three scalars' primordial perturbations. Such modes were studied in \cite{Endlich:2012pz}, where it was shown that their amplitudes is nearly frozen outside the horizon. It remains to be studied whether these modes can result in a sizable anisotropic IR background, and then imprint an anisotropic contribution to $P_\zeta$, analogously to what happens in the $f \left( \phi \right) F^2 $ model.

The discussion we have just presented is on whether an isotropic classical background can be destabilized by the random anisotropic addition of the super-horizon modes of the different fields. A different problem, strongly motivated by our results, is on whether solid inflation can lead to a isotropic and homogeneous background starting from generic initial conditions. We have shown here that solid inflation erases an initial anisotropy on a rather long timescale, $\Delta t = {\rm O } \left( \frac{1}{H \epsilon} \right)$. We understood this in terms of the fact that the medium of solid inflation must be extremely insensitive to spatial deformations. It is natural to wonder whether an analogous inefficiency will also take place for an initially inhomogeneous background. This would question the validity of solid inflation as a solution of the homogeneity and isotropy problem, in contrast to more standard models of inflation  \cite{Linde:2005ht}.

Finally, an open question already pointed out in  \cite{Endlich:2012pz} is related to the physics of reheating. It was shown in  \cite{Endlich:2012pz} that the primordial perturbation $\zeta$ is constant if reheating occurs instantaneously. It is possible that this is no longer the case for a more prolonged duration (we would not expect that the qualitative features of the perturbations will be changed in this case). This would require entering in the details of the field theory described by solid inflation, and of how it is coupled to ordinary matter, which by itself would also be an interesting direction to explore.

\vskip.25cm
\noindent{\bf Acknowledgements:}

We thank Guillermo Ballesteros and  Eiichiro Komatsu for useful discussions. 
The work of N.B. and S.M. was partially supported by the ASI/INAF Agreement 
I/072/09/0 for the Planck LFI Activity of Phase E2. N.B, S.M. and A.R.  were also supported by the PRIN 2009 project "La Ricerca di non-Gaussianit{\' a} Primordiale". The work of  M.P. was partially supported  by DOE grant DE-FG02-94ER-40823 at the University of Minnesota. MP would like to thank the University of Padova, INFN, Sezione di Padova, and the Cosmology Group at the Department of Theoretical Physics of the University of Geneva for their friendly hospitality and for partial support during his sabbatical leave.


\begin{thebibliography}{99}



\bibitem{Ade:2013uln} 
  P.~A.~R.~Ade {\it et al.}  [Planck Collaboration],
  arXiv:1303.5082 [astro-ph.CO].


\bibitem{Cheung:2007st} 
  C.~Cheung, P.~Creminelli, A.~L.~Fitzpatrick, J.~Kaplan and L.~Senatore,
  JHEP {\bf 0803}, 014 (2008)
  [arXiv:0709.0293 [hep-th]].


\bibitem{Weinberg:2008hq} 
  S.~Weinberg,
  Phys.\ Rev.\ D {\bf 77}, 123541 (2008)
  [arXiv:0804.4291 [hep-th]].


\bibitem{Pajer:2013fsa} 
  E.~Pajer and M.~Peloso,
  arXiv:1305.3557 [hep-th].




\bibitem{Linde:2005ht} 
  A.~D.~Linde,
  Contemp.\ Concepts Phys.\  {\bf 5}, 1 (1990)
  [hep-th/0503203].


\bibitem{Collins:1972tf} 
  C.~B.~Collins and S.~W.~Hawking,
  Astrophys.\ J.\  {\bf 180}, 317 (1973).




\bibitem{triad}
  M.~C.~Bento, O.~Bertolami, P.~V.~Moniz, J.~M.~Mourao and P.~M.~Sa,
  Class.\ Quant.\ Grav.\  {\bf 10}, 285 (1993)
  [gr-qc/9302034];
%
  Y.~Hosotani,
  Phys.\ Lett.\ B {\bf 147}, 44 (1984);
%
  D.~V.~Galtsov and M.~S.~Volkov,
  Phys.\ Lett.\ B {\bf 256}, 17 (1991);
%
 C.~Armendariz-Picon,
  JCAP {\bf 0407}, 007 (2004)
  [astro-ph/0405267];
%
  A.~Maleknejad and M.~M.~Sheikh-Jabbari,
  Phys.\ Lett.\ B {\bf 723}, 224 (2013)
  [arXiv:1102.1513 [hep-ph]];
%
  P.~Adshead and M.~Wyman,
  Phys.\ Rev.\ Lett.\  {\bf 108}, 261302 (2012)
  [arXiv:1202.2366 [hep-th]];
%
  K.~-i.~Maeda and K.~Yamamoto,
  Phys.\ Rev.\ D {\bf 87}, 023528 (2013)
  [arXiv:1210.4054 [astro-ph.CO]].



\bibitem{Golovnev:2008cf} 
  A.~Golovnev, V.~Mukhanov and V.~Vanchurin,
  JCAP {\bf 0806}, 009 (2008)
  [arXiv:0802.2068 [astro-ph]].

\bibitem{massive-V} 
  K.~Dimopoulos, M.~Karciauskas and J.~M.~Wagstaff,
  Phys.\ Rev.\ D {\bf 81}, 023522 (2010)
  [arXiv:0907.1838 [hep-ph]];
 %
  K.~Dimopoulos, M.~Karciauskas and J.~M.~Wagstaff,
  Phys.\ Lett.\ B {\bf 683}, 298 (2010)
  [arXiv:0909.0475 [hep-ph]];
%
  R.~Namba,
  Phys.\ Rev.\ D {\bf 86}, 083518 (2012)
  [arXiv:1207.5547 [astro-ph.CO]];
%
  J.~A.~R.~Cembranos, C.~Hallabrin, A.~L.~Maroto and S.~J.~N.~Jareno,
  Phys.\ Rev.\ D {\bf 86}, 021301 (2012)
  [arXiv:1203.6221 [astro-ph.CO]];
%
  J.~A.~R.~Cembranos, A.~L.~Maroto and S.~J.~N.~Jareno,
  Phys.\ Rev.\ D {\bf 87}, 043523 (2013)
  [arXiv:1212.3201 [astro-ph.CO]].


\bibitem{ArmendarizPicon:2007nr} 
  C.~Armendariz-Picon,
  JCAP {\bf 0709}, 014 (2007)
  [arXiv:0705.1167 [astro-ph]].

\bibitem{Endlich:2012pz} 
  S.~Endlich, A.~Nicolis and J.~Wang,
  arXiv:1210.0569 [hep-th].

\bibitem{Gruzinov:2004ty} 
  A.~Gruzinov,
  Phys.\ Rev.\ D {\bf 70}, 063518 (2004)
  [astro-ph/0404548].




\bibitem{Shiraishi:2013vja} 
  M.~Shiraishi, E.~Komatsu, M.~Peloso and N.~Barnaby,
  JCAP {\bf 1305}, 002 (2013)
  [arXiv:1302.3056 [astro-ph.CO]].




\bibitem{Babich:2004gb} 
  D.~Babich, P.~Creminelli and M.~Zaldarriaga,
  JCAP {\bf 0408}, 009 (2004)
  [astro-ph/0405356].

\bibitem{Lewis:2011au} 
  A.~Lewis,
  JCAP {\bf 1110}, 026 (2011)
  [arXiv:1107.5431 [astro-ph.CO]].




\bibitem{Barnaby:2012tk} 
  N.~Barnaby, R.~Namba and M.~Peloso,
  Phys.\ Rev.\ D {\bf 85}, 123523 (2012)
  [arXiv:1202.1469 [astro-ph.CO]].


\bibitem{Bartolo:2012sd} 
  N.~Bartolo, S.~Matarrese, M.~Peloso and A.~Ricciardone,
  Phys.\ Rev.\ D {\bf 87}, 023504 (2013)
  [arXiv:1210.3257 [astro-ph.CO]].


\bibitem{Funakoshi:2012ym} 
  H.~Funakoshi and K.~Yamamoto,
  Class.\  Quantum Grav.\  {\bf 30}, 135002 (2013)
  [arXiv:1212.2615 [astro-ph.CO]].


\bibitem{Abolhasani:2013zya} 
  A.~A.~Abolhasani, R.~Emami, J.~T.~Firouzjaee and H.~Firouzjahi,
  arXiv:1302.6986 [astro-ph.CO].


\bibitem{Biagetti:2013qqa} 
  M.~Biagetti, A.~Kehagias, E.~Morgante, H.~Perrier and A.~Riotto,
  arXiv:1304.7785 [astro-ph.CO].

\bibitem{Fujita:2013qxa} 
  T.~Fujita and S.~Yokoyama,
  arXiv:1306.2992 [astro-ph.CO].


\bibitem{Lyth:2013sha} 
  D.~H.~Lyth and M.~Karciauskas,
  JCAP {\bf 1305}, 011 (2013)
  [arXiv:1302.7304 [astro-ph.CO]].

\bibitem{Yokoyama:2008xw} 
  S.~Yokoyama and J.~Soda,
  JCAP {\bf 0808}, 005 (2008)
  [arXiv:0805.4265 [astro-ph]].
 
\bibitem{Ade:2013ydc}
  P.~A.~R.~Ade {\it et al.}  [Planck Collaboration],
  arXiv:1303.5084 [astro-ph.CO].


\bibitem{Maleknejad:2012fw} 
  A.~Maleknejad, M.~M.~Sheikh-Jabbari and J.~Soda,
  arXiv:1212.2921 [hep-th].


\bibitem{Ford:1989me} 
  L.~H.~Ford,
  Phys.\ Rev.\ D {\bf 40}, 967 (1989).
    
\bibitem{Turner:1987bw}
  M.~S.~Turner and L.~M.~Widrow,
  Phys.\ Rev.\ D {\bf 37} (1988) 2743.
  
 
 
\bibitem{Dimopoulos:2008yv} 
  K.~Dimopoulos, M.~Karciauskas, D.~H.~Lyth and Y.~Rodriguez,
  JCAP {\bf 0905}, 013 (2009)
  [arXiv:0809.1055 [astro-ph]].
 

 
\bibitem{Ackerman:2007nb} 
  L.~Ackerman, S.~M.~Carroll and M.~B.~Wise,
  Phys.\ Rev.\ D {\bf 75}, 083502 (2007)
  [Erratum-ibid.\ D {\bf 80}, 069901 (2009)]
  [astro-ph/0701357].



\bibitem{hcp}
  B.~Himmetoglu, C.~R.~Contaldi and M.~Peloso,
  Phys.\ Rev.\ Lett.\  {\bf 102}, 111301 (2009)
  [arXiv:0809.2779 [astro-ph]];
 %
  B.~Himmetoglu, C.~R.~Contaldi and M.~Peloso,
  Phys.\ Rev.\ D {\bf 79}, 063517 (2009)
  [arXiv:0812.1231 [astro-ph]];
  %
  B.~Himmetoglu, C.~R.~Contaldi and M.~Peloso,
  Phys.\ Rev.\ D {\bf 80}, 123530 (2009)
  [arXiv:0909.3524 [astro-ph.CO]].


\bibitem{Himmetoglu:2009mk} 
  B.~Himmetoglu,
  JCAP {\bf 1003}, 023 (2010)
  [arXiv:0910.3235 [astro-ph.CO]].
  
 

\bibitem{Ratra:1991bn} 
  B.~Ratra,
  Astrophys.\ J.\  {\bf 391}, L1 (1992).
             
                             
\bibitem{Martin:2007ue}
  J.~Martin and J.~'i.~Yokoyama,
  JCAP {\bf 0801} (2008) 025
  [arXiv:0711.4307 [astro-ph]].
              

\bibitem{Giovannini:2009xa} 
  M.~Giovannini,
  JCAP {\bf 1004}, 003 (2010)
  [arXiv:0911.0896 [astro-ph.CO]].
  
 
\bibitem{Demozzi:2009fu} 
  V.~Demozzi, V.~Mukhanov and H.~Rubinstein,
  JCAP {\bf 0908}, 025 (2009)
  [arXiv:0907.1030 [astro-ph.CO]].
      
\bibitem{Fujita:2012rb} 
  T.~Fujita and S.~Mukohyama,
  JCAP {\bf 1210}, 034 (2012)
  [arXiv:1205.5031 [astro-ph.CO]].

\bibitem{Ferreira:2013sqa} 
  R.~J.~Z.~Ferreira, R.~K.~Jain and M.~S.~Sloth,
  arXiv:1305.7151 [astro-ph.CO].

\bibitem{Watanabe:2009ct} 
  M.~-a.~Watanabe, S.~Kanno and J.~Soda,
  Phys.\ Rev.\ Lett.\  {\bf 102}, 191302 (2009)
  [arXiv:0902.2833 [hep-th]].
 
 \bibitem{aniso-fAA} 
  R.~Emami, H.~Firouzjahi, S.~M.~Sadegh Movahed and M.~Zarei,
  JCAP {\bf 1102}, 005 (2011)
  [arXiv:1010.5495 [astro-ph.CO]];
 %
  S.~Kanno, J.~Soda and M.~-a.~Watanabe,
  JCAP {\bf 1012}, 024 (2010)
  [arXiv:1010.5307 [hep-th]];
 %
  K.~Murata and J.~Soda,
  JCAP {\bf 1106}, 037 (2011)
  [arXiv:1103.6164 [hep-th]];
  %
    T.~Q.~Do, W.~F.~Kao and I.~-C.~Lin,
  Phys.\ Rev.\ D {\bf 83}, 123002 (2011).
   %
  T.~Q.~Do and W.~F.~Kao,
  Phys.\ Rev.\ D {\bf 84}, 123009 (2011).
  %
  K.~Yamamoto, M.~-a.~Watanabe and J.~Soda,
  Class.\ Quant.\ Grav.\  {\bf 29}, 145008 (2012)
  [arXiv:1201.5309 [hep-th]];
  %
  M.~Thorsrud, D.~F.~Mota and S.~Hervik,
  arXiv:1205.6261 [hep-th].
 
  
\bibitem{Matarrese:1984zw} 
  S.~Matarrese,
  Proc.\ Roy.\ Soc.\ Lond.\ A {\bf 401}, 53 (1985).
 
\bibitem{Arroja:2010wy} 
  F.~Arroja and M.~Sasaki,
  Phys.\ Rev.\ D {\bf 81}, 107301 (2010)
  [arXiv:1002.1376 [astro-ph.CO]].

\bibitem{Chen:2013kta} 
  X.~Chen, H.~Firouzjahi, M.~H.~Namjoo and M.~Sasaki,
  arXiv:1306.2901 [hep-th].

 
 
\bibitem{Wald:1983ky} 
  R.~M.~Wald,
  Phys.\ Rev.\ D {\bf 28}, 2118 (1983).
  
\bibitem{Maleknejad:2012as} 
  A.~Maleknejad and M.~M.~Sheikh-Jabbari,
  Phys.\ Rev.\ D {\bf 85}, 123508 (2012)
  [arXiv:1203.0219 [hep-th]].



\bibitem{R2aniso}
  J.~D.~Barrow and S.~Hervik,
  Phys.\ Rev.\ D {\bf 73}, 023007 (2006)
  [gr-qc/0511127];
%
  J.~D.~Barrow and S.~Hervik,
  Phys.\ Rev.\ D {\bf 74}, 124017 (2006)
  [gr-qc/0610013].


\bibitem{Kaloper:1991rw} 
  N.~Kaloper,
  Phys.\ Rev.\ D {\bf 44}, 2380 (1991).



\bibitem{DiGrezia:2003ug} 
  E.~Di Grezia, G.~Esposito, A.~Funel, G.~Mangano and G.~Miele,
  Phys.\ Rev.\ D {\bf 68}, 105012 (2003)
  [gr-qc/0305050].



\bibitem{Ohashi:2013mka} 
  J.~Ohashi, J.~Soda and S.~Tsujikawa,
  Phys.\ Rev.\ D {\bf 87}, 083520 (2013)
  [arXiv:1303.7340 [astro-ph.CO]].






\bibitem{Ballesteros:2012kv} 
  G.~Ballesteros and B.~Bellazzini,
  JCAP {\bf 1304}, 001 (2013)
  [arXiv:1210.1561 [hep-th]].
  
  

\bibitem{Bardeen:1980kt} 
  J.~M.~Bardeen,
  Phys.\ Rev.\ D {\bf 22}, 1882 (1980).


\bibitem{Mukhanov:1990me} 
  V.~F.~Mukhanov, H.~A.~Feldman and R.~H.~Brandenberger,
  Phys.\ Rept.\  {\bf 215}, 203 (1992).


\bibitem{Arnowitt:1962hi} 
  R.~L.~Arnowitt, S.~Deser and C.~W.~Misner,
  gr-qc/0405109.


\bibitem{Malik:2008im} 
  K.~A.~Malik and D.~Wands,
  Phys.\ Rept.\  {\bf 475}, 1 (2009)
  [arXiv:0809.4944 [astro-ph]].


\bibitem{Gumrukcuoglu:2007bx} 
  A.~E.~Gumrukcuoglu, C.~R.~Contaldi and M.~Peloso,
  JCAP {\bf 0711}, 005 (2007)
  [arXiv:0707.4179 [astro-ph]].
  
       
\bibitem{Komatsu:2010fb} 
  E.~Komatsu {\it et al.}  [WMAP Collaboration],
  Astrophys.\ J.\ Suppl.\  {\bf 192}, 18 (2011)
  [arXiv:1001.4538 [astro-ph.CO]].
    

\bibitem{Groeneboom:2009cb} 
  N.~E.~Groeneboom, L.~Ackerman, I.~K.~Wehus and H.~K.~Eriksen,
  Astrophys.\ J.\  {\bf 722}, 452 (2010)
  [arXiv:0911.0150 [astro-ph.CO]].


   
\bibitem{Hanson:2009gu}
  D.~Hanson and A.~Lewis,
  Phys.\ Rev.\ D {\bf 80} (2009) 063004
  [arXiv:0908.0963 [astro-ph.CO]].

\bibitem{Hanson:2010gu} 
  D.~Hanson, A.~Lewis and A.~Challinor,
  Phys.\ Rev.\ D {\bf 81}, 103003 (2010)
  [arXiv:1003.0198 [astro-ph.CO]].




\bibitem{Bennett:2012fp} 
  C.~L.~Bennett, D.~Larson, J.~L.~Weiland, N.~Jarosik, G.~Hinshaw, N.~Odegard, K.~M.~Smith and R.~S.~Hill {\it et al.},
  arXiv:1212.5225 [astro-ph.CO].



\bibitem{Ade:2013nlj} 
  P.~A.~R.~Ade {\it et al.}  [Planck Collaboration],
  arXiv:1303.5083 [astro-ph.CO].

\bibitem{Pullen:2010zy} 
  A.~R.~Pullen and C.~M.~Hirata,
  JCAP {\bf 1005}, 027 (2010)
  [arXiv:1003.0673 [astro-ph.CO]].

     
\bibitem{Dulaney:2010sq} 
  T.~R.~Dulaney and M.~I.~Gresham,
  Phys.\ Rev.\ D {\bf 81}, 103532 (2010)
  [arXiv:1001.2301 [astro-ph.CO]].
  
\bibitem{Gumrukcuoglu:2010yc} 
  A.~E.~Gumrukcuoglu, B.~Himmetoglu and M.~Peloso,
  Phys.\ Rev.\ D {\bf 81}, 063528 (2010)
  [arXiv:1001.4088 [astro-ph.CO]].
  
\bibitem{Watanabe:2010fh} 
  M.~-a.~Watanabe, S.~Kanno and J.~Soda,
  Prog.\ Theor.\ Phys.\  {\bf 123}, 1041 (2010)
  [arXiv:1003.0056 [astro-ph.CO]].



\bibitem{Affleck:1984fy} 
  I.~Affleck and M.~Dine,
  Nucl.\ Phys.\ B {\bf 249}, 361 (1985).



\bibitem{Lyth:2001nq} 
  D.~H.~Lyth and D.~Wands,
  Phys.\ Lett.\ B {\bf 524}, 5 (2002)
  [hep-ph/0110002].




\end{thebibliography}
\end{document}